%% file: paper7633_afterCR2.tex
\documentclass[ALICE,manyauthors]{cernphprep}

\usepackage{amsmath}
\usepackage{bm}

\usepackage[comma,square,numbers,sort&compress]{natbib}
\usepackage{color}
\usepackage{hyperref}

\usepackage{lineno}
\usepackage{multirow} 
\usepackage{ltxcmds}

\usepackage{placeins}
\usepackage{subfigure}

\usepackage[T1]{fontenc}
\usepackage{orcidlink}

\begin{document}

\begin{titlepage}
\PHyear{2023}
\PHnumber{229}      
\PHdate{10 October}  
%

\title{Femtoscopic correlations of identical charged pions and kaons in pp collisions at {\ensuremath{\pmb{\sqrt{s}}}{\bf ~=~13}~TeV} with event-shape selection}
\ShortTitle{Femtoscopic correlations of identical charged pions and kaons in pp collisions}   

\Collaboration{ALICE Collaboration\thanks{See Appendix~\ref{app:collab} for the list of collaboration members}}
\ShortAuthor{ALICE Collaboration} 

\begin{abstract}
Collective behavior has been observed in high-energy heavy-ion collisions for several decades. Collectivity is driven by the high particle multiplicities that are produced in these collisions. At the CERN Large Hadron Collider (LHC), features of collectivity have also been seen in high-multiplicity proton--proton collisions that can attain particle multiplicities comparable to peripheral Pb--Pb collisions. One of the possible signatures of collective behavior is the decrease of femtoscopic radii extracted from pion and kaon pairs emitted from high-multiplicity collisions with increasing pair transverse momentum. This decrease can be described in terms of an approximate transverse mass scaling. In the present work, femtoscopic analyses are carried out by the ALICE Collaboration on charged pion and kaon pairs produced in pp collisions at $\sqrt{s}=13$ TeV from the LHC to study possible collectivity in pp collisions. The event-shape analysis method based on transverse sphericity is used to select for spherical versus jet-like events, and the effects of this selection on the femtoscopic radii for both charged pion and kaon pairs are studied. This is the first time this selection method has been applied to charged kaon pairs. An approximate transverse-mass scaling of the radii is found in all multiplicity ranges studied when the difference in the Lorentz boost for pions and kaons is taken into account. This observation does not support the hypothesis of collective expansion of hot and dense matter that should only occur in high-multiplicity events. A possible alternate explanation of the present results is based on a scenario of common emission conditions for pions and kaons in pp collisions for the multiplicity ranges studied.
\end{abstract}
\end{titlepage}
\setcounter{page}{2}

\section{Introduction}

The manifestation of collective effects in pp (${\rm p}\overline{\rm p}$) and p--A collisions with increasing multiplicity of charged particles is intensely discussed in the literature~\cite{Alexopoulos:1992iv,Nagle:2018nvi,Adolfsson:2020dhm,ALICE:2022wpn}.
Surprisingly, these small colliding systems exhibit several signatures
attributed to the formation of a strongly-interacting quark--gluon plasma in
heavy-ion collisions, such as long-range ridge-like structures~\cite{CMS:2010ifv,ALICE:2012eyl,ATLAS:2015hzw} and strangeness enhancement~\cite{ALICE:2016fzo}. A full understanding of the mechanisms leading to collective effects observed in pp collisions at large multiplicity has not yet been achieved. For this reason, it is important to experimentally investigate the properties of small systems with the aim to discriminate between different theoretical models.
Namely, the hydrodynamic models present the ``heavy-ion view'' of pp collisions, e.g.\ Ref.~\cite{Pierog:2013ria}, while string models are the ``high-energy view'', e.g.\ models including interactions between strings~\cite{Bierlich:2018xfw,Fischer:2016zzs}.

The femtoscopy technique, which studies the final-state hadron--hadron interactions via their momentum correlations, is an effective tool for the extraction of the space--time characteristics of particle production processes, in particular the radii of the emitting source and the decoupling time. This method has already been employed in the past to study high-energy hadron--hadron~\cite{Kittel:2001zw,Alexander:2003ug} and heavy-ion collisions~\cite{Podgoretsky:1989bp,Lednicky:2005af} using quantum statistical (QS) correlations and/or final-state interactions (FSI) of particles emitted with small relative momenta.

The characteristic feature of femtoscopy in heavy-ion collisions is the decrease of the source sizes for pairs of particles (with masses $m$ and transverse momenta ${\bf p}_{\mathrm{T,1}}$ and ${\bf p}_{\mathrm{T,2}}$) with increasing pair transverse momentum \mbox{${\it k_{\mathrm T}}=|{\bf p}_{\mathrm{T,1}} + {\bf p}_{\mathrm{T,2}}|/2$} or transverse mass
\mbox{$\it{ m_{\mathrm T}}  = \sqrt{{\it k_\mathrm{T}}^2+m^{2}}$},
see e.g.\ recent results for pions from the BNL Relativistic Heavy Ion Collider (RHIC)~\cite{STAR:2009fks,STAR:2004qya,PHOBOS:2004sjz} and the CERN Large Hadron Collider (LHC)~\cite{ALICE:2011dyt,ALICE:2015tra}. It was explained in Ref.~\cite{Akkelin:1995gh} that femtoscopy measurements do not probe the whole volume in the case of an expanding emitting source, but instead the region from which the particles with similar momenta are emitted, the so called ``homogeneity volume''. This region is smaller than the total volume occupied by the system and decreases with \mbox{$\it{k_{\mathrm T}}$}
(\mbox{$\it{m_{\mathrm T}}$}).
Several theoretical models based on the hydrodynamic approach successfully describe pion femtoscopy measurements,
e.g.\ Refs.~\cite{Karpenko:2009wf,Shapoval:2020nec,Kisiel:2005hn,Bozek:2010wt,Kisiel:2014upa}.
It is expected that these models should describe the femtoscopy measurements for kaons and for heavier particles also, particularly the \mbox{$\it{m_{\mathrm T}}$} and multiplicity dependencies of radii. In Ref.~\cite{Makhlin:1987gm}, it was shown that for the particular case of small transverse flow the hydrodynamics leads to the same \mbox{$\it{m_{\mathrm T}}$} behavior of the longitudinal radii ($R_{\rm long}$) for pions and kaons. It means that the thermal freeze-out occurs simultaneously and that these two particle species are subject to the same velocity boost from collective flow. Modern calculations made within the 3+1-dimensional (3+1D) hydrodynamic model THERMINATOR-2~\cite{Kisiel:2005hn} at LHC energies demonstrate the approximate \mbox{$\it{m_{\mathrm T}}$} scaling of the three-dimensional radii for pions, kaons, and protons~\cite{Kisiel:2014upa} when the radii versus \mbox{$\it{m_{\mathrm T}}$} fall with some degree of accuracy on one curve. The authors of Ref.~\cite{Kisiel:2014upa} also investigated one-dimensional radii ($R_{\rm inv}$) in the pair reference frame (PRF) for the case of lack of available experimental data. They have verified that the violation of this scaling in the case when the three-dimensional scaling is presented in the model has a trivial kinematic origin. It is possible to take it into account and restore the $m_{\rm T}$ scaling for $R_{\rm inv}$ for pions, kaons, and protons. The calculations performed within the Hydro-Kinetic Model, including not only a hydrodynamic phase but also the hadronic rescattering stage, predicted violation of this scaling between pions and kaons at LHC energies~\cite{Shapoval:2014wya,Sinyukov:2015kga}, mainly due to the rescatterings in the hadronic phase. ALICE results for Pb--Pb collisions at $\sqrt{s_{\rm{NN}}}=2.76$~TeV~\cite{ALICE:2017iga} have shown that the
\mbox{$\it{m_{\mathrm T}}$} scaling expected by pure hydrodynamical scenarios is broken. The comparison of these two different evolution scenarios demonstrate the importance of the \mbox{$\it{m_{\mathrm T}}$} scaling study.

At the LHC, the femtoscopic sizes range for various colliding systems from 2--7~fm for Pb--Pb to 1--2~fm for pp and p--Pb collisions, which opens a possible access to different energy densities of the system created during such collisions and probably helps understand the conditions required for the QGP formation. The multiplicity and \mbox{$\it{m_{\mathrm T}}$} dependencies were studied for pions and kaons by ALICE for Pb--Pb collisions (e.g.\ Refs.~\cite{ALICE:2015tra,ALICE:2017iga}) and for pp and p--Pb collisions (e.g.\ Refs.~\cite{ALICE:2011kmy,ALICE:2015hav,ALICE:2012yyu,Abelev:2012sq,ALICE:2019kno,Adam:2015vja}).
In pp collisions, it was observed for both types of pairs that for higher charged-particle multiplicity ranges 
the measured size of the source decreased with increasing $\it{m_{\mathrm T}}$, similarly to the trend seen in heavy-ion collisions.
Instead, at low charged-particle multiplicities, the measured source radii increased with increasing $\it{m_{\mathrm T}}$~\cite{ALICE:2011kmy,ALICE:2012yyu,Abelev:2012sq}.
Unfortunately, there are almost no theoretical models for pp collisions which include the space--time coordinates and can be used to describe femtoscopic observables. Attempts to describe the behavior of femtoscopic radii in pp collisions at the LHC from the hydrodynamic point of view were performed in e.g.\ Refs.~\cite{Werner:2010ny,Plumberg:2020jod}, from the view of the uncertainty principle in Ref.~\cite{Shapoval:2013jca}, and from the view of the string models in Ref.~\cite{Nilsson:2011xg}
using the Lund hadronization scheme which automatically introduces the space--momentum correlations, similar to the correlations in hydrodynamic models, where they arise due to transverse collective flow.

The study of femtoscopic correlations in pp collisions is more challenging than in A--A collisions due to the strong non-femtoscopic contributions, i.e.\ correlations due to multi-body resonance decays, mini-jets, and energy-momentum conservation. Typically, a baseline distribution function is constructed to remove these
non-femtoscopic effects, after which the hadron--hadron correlations due to pure
QS and FSI can be studied. There are various methods to exclude them: (1)~the ``double ratio'' technique, dividing the experimental correlation function by the baseline extracted from Monte Carlo simulations~\cite{ALICE:2011kmy}; (2)~the ``cluster subtraction'' technique, using the opposite-sign pair (e.g.~$\pi^+\pi^-$) distributions as a baseline~\cite{CMS:2019fur}; (3)~a ``hybrid method'' between (1) and (2), as described in Refs.~\cite{ATLAS:2017shk,CMS:2019fur}); and (4) the three-particle cumulant method~\cite{Abelev:2014pja}, which significantly suppresses the mini-jet related contributions and can be used as an alternative to the study of two-particle correlations.

A method to suppress, in particular, mini-jet contributions in two-particle correlation functions was suggested in Ref.~\cite{ALICE:2019bdw} and it is based on applying event-shape selections. It was shown that it is possible to differentiate between jet-like and spherical event topologies using a global characteristic of the event such the transverse sphericity~\cite{Abelev:2012sk,Banfi:2004nk} and the transverse spherocity~\cite{Ortiz:2015ttf,ALICE:2019dfi}. It was observed for the first time in Ref.~\cite{ALICE:2019bdw} that the pion radii for jet-like events are smaller than the source radii for spherical events. In jet-like events, the radii dependence on multiplicity is such that they increase with increasing $k_{\rm T}$ in the lowest multiplicity interval and decrease with $k_{\rm T}$ for the highest multiplicities. There are no multiplicity dependences for these events. The radii for spherical events show an increase in system size with increasing multiplicity. They do not show any visible trend with ${\it k_{\mathrm T}}$, which differs from the results of sphericity-integrated (minimum-bias) pion and kaon analyses in Refs.~\cite{ALICE:2011kmy,Abelev:2012sq}, where the obtained radii decrease with increasing ${\it k_{\mathrm T}}$.
This different behavior suggests~\cite{ALICE:2019bdw} that the lower part of the transverse sphericity spectrum contributes to the observed slope in minimum bias (MB) pp collisions.

In this work, the femtoscopic correlations of identical charged pions and kaons are investigated in pp collisions at $\sqrt{s}=13$~TeV. The purpose of this analysis is to study the transverse-momentum and multiplicity dependence of pion and kaon femtoscopic radii separately for jet-like and spherical events. The influence of the sphericity selections on the kaon femtoscopic radii is studied for the first time.

The article is organized as follows. Section~2 briefly describes the ALICE experimental setup. Section~3 presents the event selection criteria. Section~4 introduces the definition of sphericity, and discusses the analysis methods to extract the femtoscopic correlation function for pion and kaon pairs and the estimation of the systematic uncertainties. The extracted femtoscopic parameters are shown and discussed in Section~5. The results are summarized in Section~6.

\section{\label{Exp} Experimental setup}
A detailed description of the ALICE detector and its performance can be found in Refs.~\cite{ALICE:2008ngc,Abelev:2014ffa}. In the present analysis, the information from the Inner Tracking System (ITS)~\cite{ALICE:2010tia}, the Time Projection Chamber (TPC)~\cite{Alme:2010ke}, the Time-Of-Flight (TOF)~\cite{Akindinov:2013tea}, and the V0~\cite{Abbas:2013taa} detectors are used.

The V0 detector is used for triggering on collision events. It is composed of two small-angle scintillator arrays, located at 340~cm and 90~cm from the nominal interaction point along the beam line and covering $2.8 < \eta < 5.1$ (V0A) and $-3.7 < \eta < -1.7$ (V0C), respectively. The events are selected with the MB trigger, which requires simultaneous signals in both parts of the V0 detector in coincidence with two beam bunches crossing in the ALICE interaction region.
The rejection of pile-up events is performed by using the vertexing capabilities of the Silicon Pixel Detector (SPD)~\cite{Abelev:2014ffa}, which forms the two innermost layers of the ITS. Events with multiple vertices identified with the SPD (in-bunch pile-up) are removed from the analysis. Pile-up events from different bunch crossings are rejected by requiring the tracks to have hits in the SPD. The remaining leftover pile-up is negligible in the present analysis.

Charged particles are reconstructed with the central barrel ITS and TPC detectors placed inside a solenoidal magnet providing a uniform 0.5~T field parallel to the beam direction. The primary vertex is reconstructed using the ITS. This detector is a silicon tracker with six layers of silicon sensors covering the pseudorapidity range $|\eta| < 0.9$~\cite{ALICE:2010tia}. The TPC is the main tracking detector in ALICE, which measures the ionization energy loss of particles. The chamber is divided into two halves by a central electrode. The end caps on either side are composed of 18 sectors (covering the full azimuthal angle) with 159 pad rows placed radially in each sector. The TPC covers an acceptance of $|\eta| < 0.9$ for tracks which reach the outer radius of the detector.

Particle identification (PID) for reconstructed tracks is carried out using the TPC together with the TOF~\cite{Akindinov:2013tea} detectors. The TOF is a cylindrical detector consisting of 18 azimuthal sectors divided into five modules along the beam axis with active element multi-gap resistive plate chambers. Pions and kaons were identified using the TPC and TOF detectors. The deviation of the specific energy loss  (d$E$/d$x$) measured in the TPC from the one calculated using the Bethe--Bloch parametrization was required to be within a certain number of standard deviations ($n_{\sigma_{\mathrm{TPC}}}$). A similar $n_{\sigma_{\mathrm{TOF}}}$ method was applied for the particle identification in the TOF. The deviation is computed between the measured time of flight and the one calculated for a given particle path length, momentum, and mass.

\section{\label{Data} Data selection}

The data samples used in this work were recorded by ALICE in 2016--2018 during the LHC Run~2 period at $\sqrt{s} = 13$~TeV. After application of all selection criteria, about 10$^{9}$~minimum bias events were analyzed.

Events were accepted if they had the collision vertex position measured along the beam line within $\pm10$~cm from the nominal interaction point. The charged particle tracks were required to be reconstructed with the ITS and TPC detectors with a $\chi^{2}$ per number of degrees of freedom ($\chi^{2}/{\rm NDF})$~smaller than 4.0, and each track segment was reconstructed from at least 70 out of the 159 possible space points. The distance of closest approach (DCA) to the primary vertex was required to be smaller than 0.3~cm in both the transverse plane and the longitudinal direction.

Femtoscopic correlation functions of identical particles are sensitive to two-track reconstruction effects because the particles of interest have close momenta and close trajectories. Two kinds of two-track effects, splitting and merging, were studied. The splitting of the tracks means that one track is reconstructed as two. The track merging means that two different tracks are reconstructed as one. To remove these effects, the distance between the tracks of two particles was calculated at up to nine points throughout the TPC volume (every 20~cm, from 85~cm to 245~cm in the radial direction) and then averaged. It was required that the particles for each pair had an average TPC separation of at least 3~cm.

Pions and kaons were selected in the pseudorapidity $|\eta| < 0.8$ range.
For pions, the transverse momentum $0.15 < p_{\rm T} < 4.0$~GeV$/c$ range was used.
The pion selection criteria are presented in Table~\ref{tab:PiPicuts}.
The pion purity is about 99\% for track momenta $p <2.0$~GeV$/c$, while, for the $2 < p_{\rm T} < 4$ GeV$/c$ interval, it decreases to 80\% due to an increasing contamination from kaons.

\begin{table}
\centering
\caption{Charged pion selection criteria.}
\begin{tabular}{ll}
  \hline\hline
    Selection criterion & Value \\ \hline
    $p_{\rm T}$  & $0.15<p_{\rm T}<4.0$~GeV$/c$ \\ \hline
    $|\eta|$ & $< 0.8$ \\ \hline
    $\rm DCA_{\rm transverse}$ & $< 0.3$~cm \\ \hline
    $\rm DCA_{\rm longitudinal}$ & $< 0.3$~cm \\ \hline
    $n_{\sigma_{\rm TPC}}$ & $< 3$ (for $p < 0.5$~GeV$/c$) \\ \hline
    $\sqrt{n_{\sigma_{\mathrm{TPC}}^{2}}+n_{\sigma_{\mathrm{TOF}}}^{2}}^{{\phantom{1}}^{\phantom{1}}}$ & $< 3$ (for $0.5 < p < 4.0$~GeV$/c$) \\ \hline
    Number of track points in TPC & $\geq$~70 \\ \hline\hline
\end{tabular}
\label{tab:PiPicuts}
\end{table}
The selection criteria for the kaons are reported in Table~\ref{tab:KKcuts}.
In order to avoid strong contamination from pions, narrower momentum ranges were used for kaons, namely $0.15 < p_{\rm T} < 1.5$~GeV$/c$. The dominant contamination for charged kaons is from ${\rm e}^{\pm}$ in the particle momentum range $0.4< p < 0.5$~GeV$/c$, resulting in a kaon purity of approximately 90\%.
Outside this range, the kaon purity is about 99\%.

\begin{table}
\centering
\caption{Charged kaon selection criteria.}
\begin{tabular}{ll}
  \hline\hline
    Selection criterion & Value \\ \hline
    $p_{\rm T}$  & $0.15<p_{\rm T}<1.5$~GeV$/c$ \\ \hline
    $|\eta|$ & $< 0.8$ \\ \hline
    $\rm DCA_{\rm transverse}$ & $< 0.3$~cm \\ \hline
    $\rm DCA_{\rm longitudinal}$ & $< 0.3$~cm \\ \hline
     & $< 2$ (for $ 0.15 < p < 0.4$~GeV$/c$)\\
    $n_{\sigma_{\rm TPC}}$ & $< 1$ (for $ 0.4 < p < 0.45$~GeV$/c$) \\
     & $< 2$ (for $0.45 < p < 1.5$~GeV$/c$) \\ \hline
     & $< 2$ (for $0.5 < p < 0.8$~GeV$/c$) \\ 
    $n_{\sigma_{\rm TOF}}$ & $< 1.5$ (for $0.8 < p < 1.0$~GeV$/c$) \\
     & $< 1.0$ (for $1.0 < p < 1.5$~GeV$/c$) \\ \hline
    Number of track points in TPC  & $\geq$~70 \\ \hline\hline
\end{tabular}
\label{tab:KKcuts}
\end{table}

\section{\label{CFan}Analysis technique}
Pions and kaons were selected in the same raw charged-particle multiplicity intervals $N_{\rm trk}$ of (1--18), (19--30), and ($>30$) in order to compare the obtained results in the same multiplicity conditions. The sphericity calculations (see Section~\ref{Sphericity}) require at least three tracks with \mbox{$p_{\rm T}>0.5$}~GeV$/c$.
Therefore, the lowest multiplicity interval is (1--18) if the sphericity calculation is not performed, while it is (3--18) when the sphericity is calculated. From now on, the lowest multiplicity interval will be denoted as (1--18) for both cases. The multiplicity for primary charged tracks~\cite{ALICE:2017hcy} was estimated using the combined reference multiplicity estimator (SPD tracklets and tracks reconstructed in the ITS and the TPC) in the \mbox{$|\eta| < 0.8$} range. The SPD tracklets are track segments built by associating pairs of hits in the two SPD layers. For each raw multiplicity interval, the average charged-particle pseudorapidity density
$\langle{\rm d}N_{\rm ch}/{\rm d}\eta\rangle$ was obtained by converting the measured event multiplicities using Monte Carlo simulations with the PYTHIA 8.2 event generator~\cite{Sjostrand:2014zea} (with the Monash tune~\cite{Skands:2014pea})
and the GEANT3 package~\cite{Brun:1994aa} for the transport of the generated particles through the ALICE detector. The intervals and their corresponding $\langle{\rm d}N_{\rm ch}/{\rm d} \eta\rangle$, for both cases
with and without sphericity event selections ($S_{\rm T}$, defined in the next section), are shown in Table~\ref{tab:dNdeta}. The systematic uncertainties for these densities were evaluated as the difference between their magnitudes when taking into account, or not, the detector efficiency correction using the Monte Carlo simulation mentioned above. The estimated value for these uncertainties is about 5~$\%$.
\begin{table}
\centering
\caption{Raw charged-particle multiplicity ($N_{\rm trk}$) intervals and corresponding average $\langle{\rm d}N_{\rm ch}/{\rm d}\eta\rangle$ calculated from corrected multiplicity distributions in the $|\eta| < 0.8$ range.
The values are quoted with their systematic uncertainties, the statistical uncertainties are negligible.}
\begin{tabular}{cccc}
\hline\hline
 $N_{\rm trk}$ &   $\langle {\rm d}N_{\rm ch}/{\rm d} \eta\rangle_{S_{\rm T}>0.7}$   & $\langle{\rm d}N_{\rm ch}/{\rm d}\eta\rangle_{S_{\rm T}<0.3}$    &  $\langle{\rm d}N_{\rm ch}/{\rm d}\eta\rangle$      \\
\hline
3(1)--18&   7.8 $\pm$ 0.4   &  6.2  $\pm$ 0.3    &  5.1  $\pm$ 0.2   \\
19--30  &   15.0  $\pm$ 0.7 &  13.5  $\pm$ 0.6   &  14.3  $\pm$ 0.7   \\
$>30$    &   25.4  $\pm$ 1.0 &  21.8  $\pm$ 0.9  &  24.7   $\pm$ 0.1   \\
\hline\hline
\end{tabular}
\label{tab:dNdeta}
\end{table}

\subsection{Transverse sphericity}\label{Sphericity}
In collider experiments, the study of the shape of the emitting source is often
performed in the transverse $x$--$y$ plane in order to avoid distortions related to the Lorentz boost in the beam direction along the $z$ axis~\cite{Banfi:2004nk}. Following this principle, the transverse sphericity ($S_{\rm T}$) was used to study event characteristics in pp collisions at the LHC by the ALICE Collaboration~\cite{Abelev:2012sk}.
The transverse sphericity is defined as
\begin{equation}
S_{\rm T}=\frac{ 2 \min (\lambda_{1}, \lambda_{2})}{\lambda_{1}+ \lambda_{2}},
\label{eq:Sphericity}
\end{equation}
where $\lambda_{1}$ and $\lambda_{2}$ are the eigenvalues of the matrix of transverse particle momenta,
\begin{equation}
S_{\rm T}=\frac{1}{\sum_{i}p_{\rm T}^{i}}\sum_{i}\frac{1}{p_{\rm T}^{i}} \left(\begin{array}{cc} (p_{\rm x}^{i})^{2} & p_{\rm x}^{i} \, p_{\rm y}^{i} \\  p_{\rm x}^{i} \, p_{\rm y}^{i}  & (p_{\rm y}^{i})^{2} \end{array}\right),
\label{eq:Sphericity1}
\end{equation}
with $p_{\rm x}^i$ and $p_{\rm y}^i$ being the components of the transverse momentum for $i$-th particle $\vec{p}_{\rm T}^i$. The transverse sphericity takes values in the (0--1) range. By definition of sphericity, in case of $S_{\rm T}\rightarrow0$, the emitting source is a strongly elongated ellipse, while $S_{\rm T}\rightarrow1$ corresponds to a nearly isotropic source in momentum or coordinate space.
To ensure good resolution of the transverse sphericity calculation, only events with more than two primary tracks in \mbox{$|\eta| < 0.8 $} and \mbox{$p_{\rm T} > 0.5$}~GeV$/c$ were selected~\cite{Abelev:2012sk}. Following Ref.~\cite{ALICE:2019bdw}, \mbox{$S_{\rm T} > 0.7$} was used in this analysis to select spherical events. It is expected that, for these events, the multiple soft particle production processes dominates. For jet-like events with $S_{\rm T}<0.3$, hard processes such as jets and mini-jets become dominant.

\begin{figure}[!ht]
\begin{center}
\includegraphics[width=0.6\textwidth]{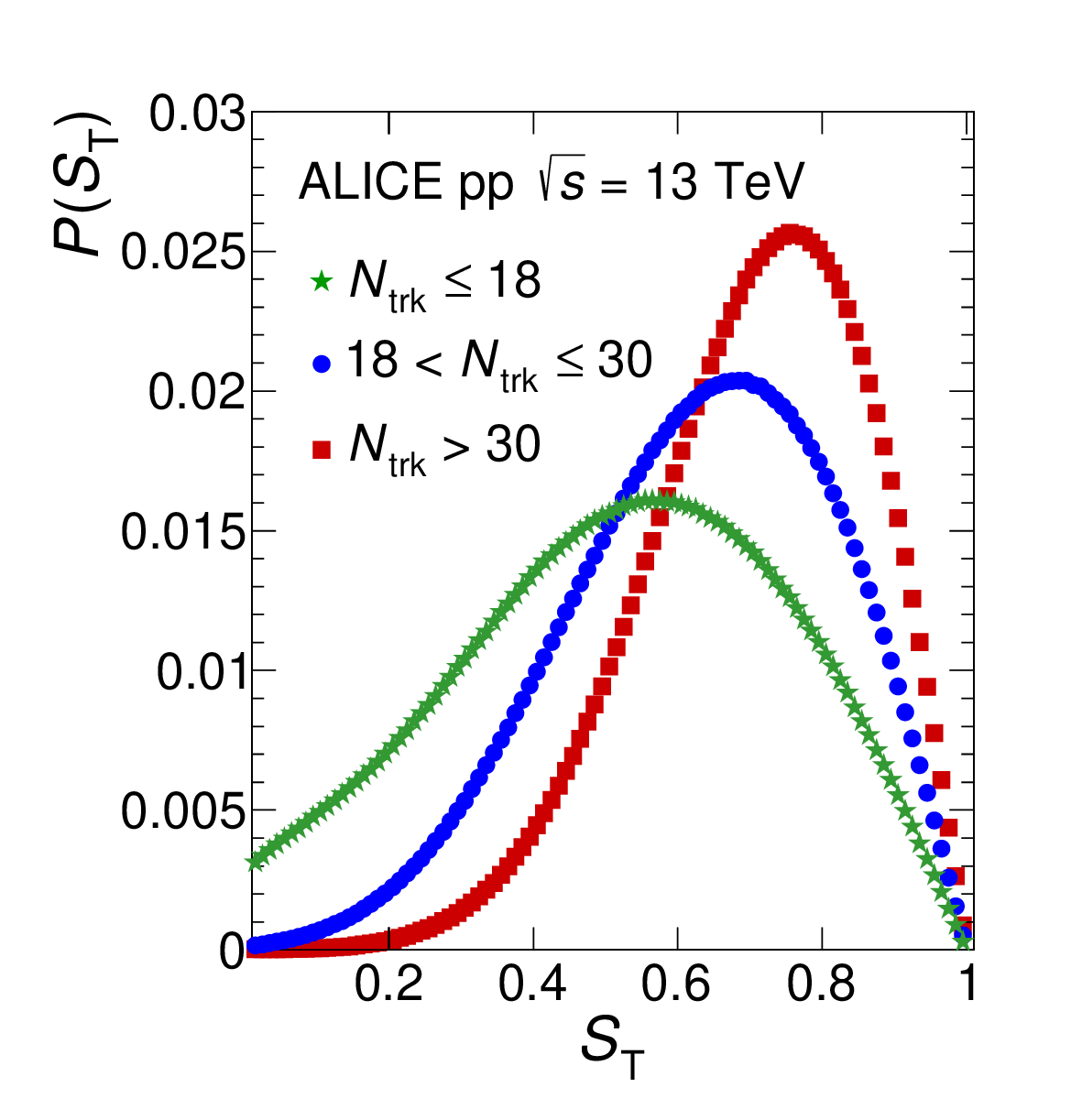}
\caption{The experimental probability $P$ of having events of different transverse sphericity $S_{\rm T}$ in the given raw multiplicity $N_{\rm trk}$ intervals: (1--18), (19--30), and ($>30$). There are no corrections applied for the efficiency of the sphericity selection. Only statistical uncertainties are shown and are smaller than the marker size.}
\label{fig:Sphericity}
\end{center}
\end{figure}

Figure~\ref{fig:Sphericity} shows the experimental probability of having events with different transverse sphericity for the given raw multiplicity intervals: (1--18), (19--30), and ($>30$). Sphericity is correlated with multiplicity, so the number of events with small $S_{\rm T}$ values is higher in the lowest multiplicity interval, while the larger multiplicity intervals contain more events with large sphericity values.

\begin{figure}[!ht]
\begin{center}
\includegraphics[width=0.6\textwidth]{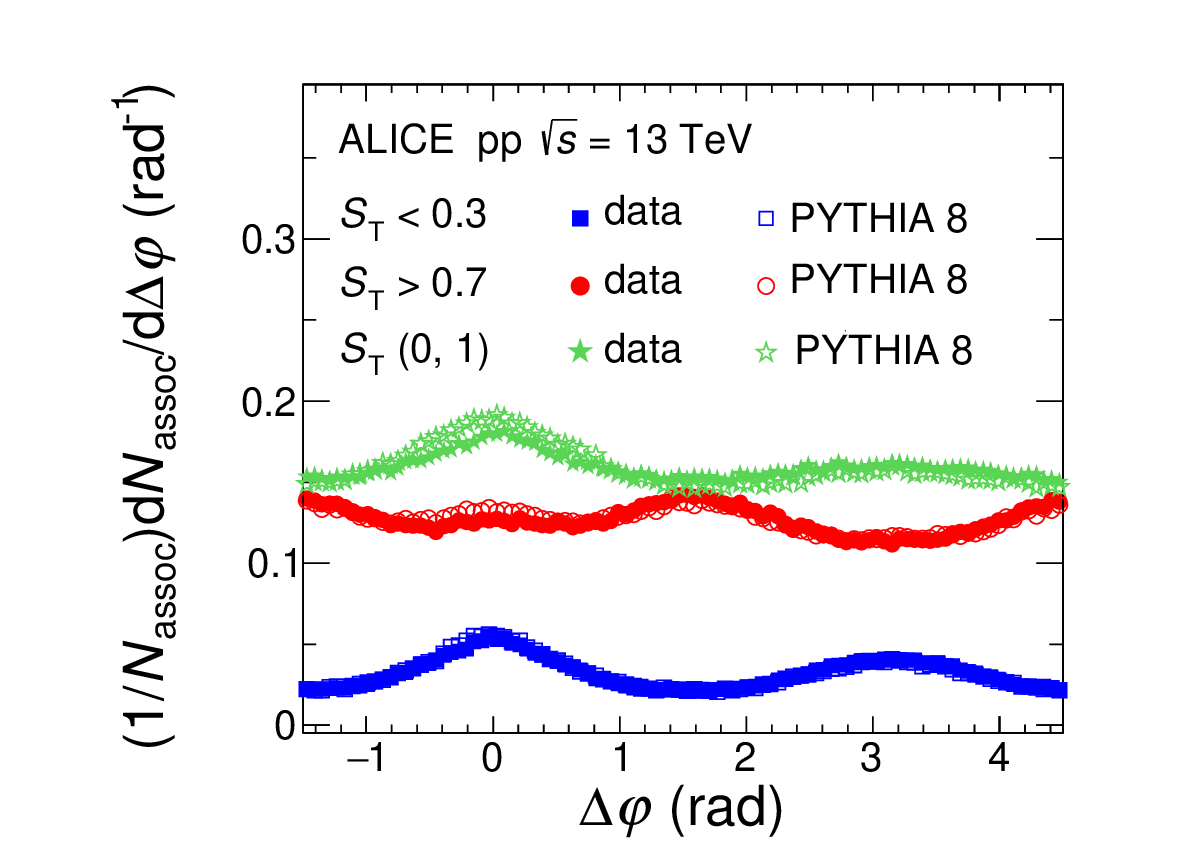}
\caption{
The pion raw experimental distribution of the azimuthal angle difference $\Delta\varphi$ between the trigger and the associated particles for $S_{\rm T}> 0.7$ (red circles), $S_{\rm T}< 0.3$ (blue squares), and $S_{\rm T}$ (0,1) (green stars) compared with MC PYTHIA 8 calculations shown with the corresponding open markers. The calculations include particle transport through the ALICE detector using the GEANT 3 transport package. The statistical uncertainties are smaller than the marker size.}
\label{fig:DPhiSt0703}
\end{center}
\end{figure}

The difference between spherical events, jet-like events, and events without sphericity selection can be clearly seen in Fig.~\ref{fig:DPhiSt0703}, which presents the experimental and Monte Carlo distributions of the azimuthal angle difference $\Delta\varphi$ between the trigger and the associated particles, where the trigger particle is the particle with the largest $p_{\rm T}$ in the event, \mbox{$p_{\rm T}^{\rm trig} > 0.5$}~GeV$/c$, \mbox{$p_{\rm T}^{\rm assoc} < p_{\rm T}^{\rm trig}$}. All distributions are normalized by the number of associated particles: $N_{\rm{assoc}}(S_{\rm T}(0,1)) = N_{\rm{assoc}}(S_{\rm T}<0.3)+N_{\rm{assoc}}(S_{\rm T}>0.7)+N_{\rm{assoc}}(S_{\rm T}(0.3,0.7))$.

The Monte Carlo simulations, with PYTHIA 8 as event generator and GEANT 3 for the simulation of the detector and the propagation of particles through the detector material, describe the $\Delta\varphi$ distributions reasonably well for all sphericity selections. The jet-like events demonstrate a strong anisotropic structure. The peak at $\Delta\varphi\approx 0$ corresponds to correlations within the jet determined by the trigger particle. The peak at $\Delta\varphi \to \pi$ corresponds to the correlation of the particle associated with the jet moving in the opposite direction. Both peaks are absent for the sphericity selection \mbox{$S_{\rm T}> 0.7$}. The $\Delta\varphi$ distribution shows some specific features with respect to the distribution without sphericity selection, which are reasonably well described by PYTHIA 8. The spherical events are much more isotropic than the jet-like ones, and the jet structures are suppressed.

\subsection{Correlation functions}\label{CFs}
The particle source created in hadronic or nuclear collisions is usually investigated using momentum correlations of two or more emitted particles. This analysis studies two-particle correlations. The observable of interest is the correlation function (CF) defined as
\begin{equation}
C({\bf p}_1,{\bf p}_2)=A({\bf p}_1,{\bf p}_2)/B({\bf p}_1,{\bf p}_2),
\label{eq:CF}
\end{equation}
where $A({\bf p}_1,{\bf p}_2)$ is the two-particle momentum distribution in the given event, and $B({\bf p}_1,{\bf p}_2)$ is a reference distribution~\cite{Kopylov:1974th}. The former includes information on the source as well as on the FSI of the emitted hadrons and/or QS effects. The latter is constructed by mixing particles emitted in two different collisions to avoid any influence of pair correlation. In the present work, the reference distribution is constructed by mixing ten events with similar multiplicity and of close sphericities. It is also required that events in a mixed event pool have their vertex positions within 2~cm from each other along the beam direction.

Due to the experimental limitation in the number of pairs, the CF is commonly defined in terms of a single kinematic variable instead of using the particle momentum vectors (see Eq.~(\ref{eq:CF})). In the following, the Lorentz-invariant $q_{\rm inv}=\sqrt{{\bf q}^2 - q_0^2}$ is used, where ${\bf q}={\bf p}_\mathrm{1}-{\bf p}_\mathrm{2}$ is the pair momentum difference and $q_0=E_1-E_2$ is the energy component difference.

The measured correlation functions are corrected using the so called double-ratio technique for the MC simulated ones $C_{\rm MC} (q_\mathrm{inv})$. This procedure assumes that the signal and the background are factorized. The corrected correlation function can be written as
\begin{equation}
C_{\rm corr}(q_\mathrm{inv})=\frac{C_{\rm data}(q_\mathrm{inv})}{C_{\rm MC}(q_\mathrm{inv})}.
\label{eq:MCcorr}
\end{equation}

Generally, the CF includes several effects, such as femtoscopic effects (QS+Coulomb in the case of $\pi^\pm\pi^\pm$ and K$^\pm$K$^\pm$ correlations), mini-jet contributions at low $q_{\rm inv}$, and long-range correlations due to energy--momentum conservation at high $q_{\rm inv}$. The latter are present in the same-event pair relative-momentum distribution and absent in the mixed-event distribution determining the CFs. If spherical events are selected with $S_{\rm T}>0.7$, contributions of mini-jets are strongly suppressed. However, there is still influence of long-range correlations due to the conservation laws. Therefore, to correct the experimental CFs for these long-range effects in spherical events, a MC model which correctly describes the shape of the experimental CFs at large $q_{\rm inv}$ are used. The experimental CFs can be divided by the MC ones (Eq.~(\ref{eq:MCcorr})), and the resulting CFs are considered to contain only femtoscopic effects, which can be fitted with a function including QS and Coulomb interaction.
For pp collisions at $\sqrt{s}=13$~TeV, the PYTHIA 8~\cite{Sjostrand:2014zea} MC model gives the best description of the experimental function outside the low $q_{\rm inv}$ region.
For jet-like ($S_{\rm T}<0.3$) events, there is a large contribution of mini-jets at low $q_{\rm inv}$ in addition to the long-range correlations at high pair relative momentum. They can also be corrected using PYTHIA~8 calculations in order to consider femtoscopic correlations only.

The analysis was performed separately for positively and negatively charged pions and kaons at two magnetic field polarities, after which the two-particle correlations were combined using their statistical uncertainties as weights.

The analysis for pions and kaons was performed in the same three multiplicity intervals. For pions, five pair transverse momentum $k_{\mathrm T}$ intervals were used: (0.15--0.3), (0.3--0.5), (0.5--0.7), (0.7--0.9), and (0.9--1.2)~GeV$/c$.
The analysis for kaons was performed in two $k_{\mathrm T}$ intervals: (0.15--0.5), (0.5--1.2)~GeV$/c$.

\subsubsection{Pion correlation functions}\label{CF_pi}
\begin{figure}[h]
\begin{center}
\includegraphics[width=1.0\textwidth]{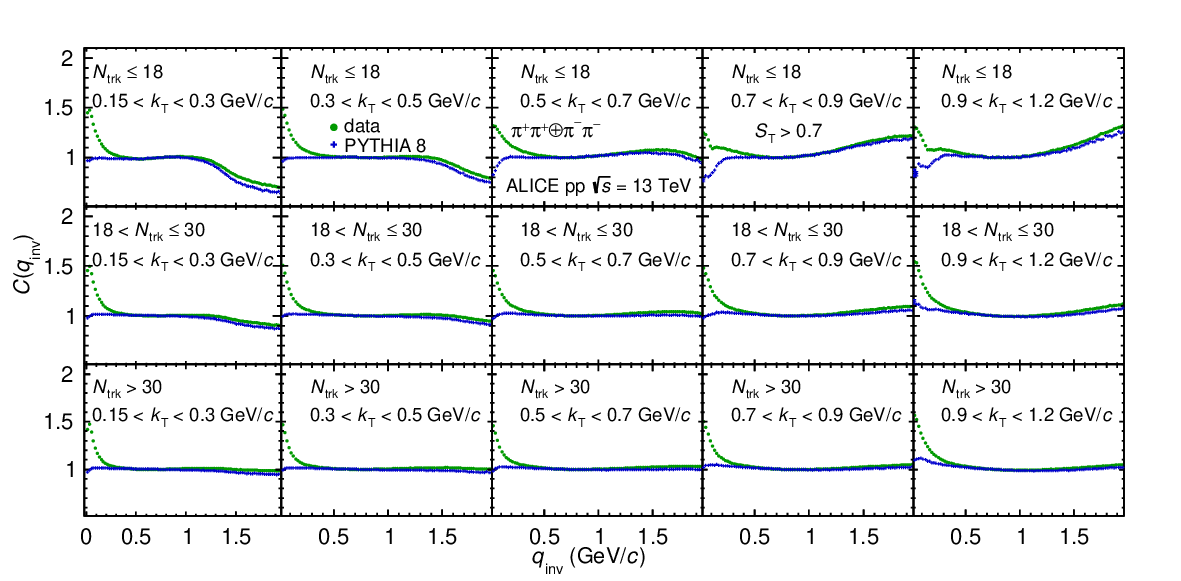}
\caption{The $\pi^\pm\pi^\pm$ experimental correlation functions (green solid circles) as function of the invariant pair relative momentum $q_\mathrm{inv}$ in pp collisions at $\sqrt{s}=13$~TeV for the raw multiplicity $N_{\rm trk}$ intervals of (1--18), (19--30), and ($>30$) in the (0.15--0.3), (0.3--0.5), (0.5--0.7), (0.7--0.9), and (0.9--1.2)~GeV$/c$ $k_{\rm T}$ intervals. A sphericity selection of $S_{\rm T} > 0.7$ is applied. The data are compared with PYTHIA 8 calculations, shown by blue crosses. The error bars represent the statistical uncertainties, while the systematic uncertainties are negligible.}
\label{fig:CFsSp07}
\end{center}
\end{figure}

Figure~\ref{fig:CFsSp07} shows the pion experimental CF (green solid circles) for events with $S_{\rm T} > 0.7$ in pp collisions at $\sqrt{s}=13$~TeV.
The PYTHIA~8 (Monash) model calculations including the ALICE detector response were used to describe non-femtoscopic effects and are also shown in the figure (blue crosses). All distributions are normalized to unity in the \mbox{$0.7 < q_{\rm inv} < 0.8$}~GeV$/c$ range, which is well outside the QS and Coulomb FSI region (\mbox{$q_{\rm inv} \lesssim 0.4$}~GeV$/c$) and before the noticeable large $q_{\rm inv}$ slope associated with energy and momentum conservation. The CFs shown in Fig.~\ref{fig:CFsSp07} are flat for the low multiplicity intervals in the region $0.5 < q_{\rm inv} < 1.0$~GeV$/c$. For the two highest $k_{\rm T}$ intervals, some slope of baseline appears. At \mbox{$q_{\rm inv} > 1.0$}~GeV$/c$, the aforementioned kinematic effects are present especially for the lowest multiplicity intervals.

The CFs decrease at \mbox{$q_{\rm inv} \to 0$} for the PYTHIA 8 calculations for the lowest multiplicity interval ($N_{\rm trk}\leq$18) in the (0.5--0.7), (0.7--0.9), and (0.9--1.2)~GeV$/c$ $k_{\rm T}$ intervals. The occurrence of such minima is related to the three-track requirement, necessary for the transverse sphericity calculation~\cite{Abelev:2012sk}. Indeed, the number of available events with three tracks with \mbox{$p_{\rm T} >0.5$}~GeV$/c$ decreases at small $q_{\rm inv}$.

\begin{figure}[h]
\begin{center}
\includegraphics[width=1.0\textwidth]{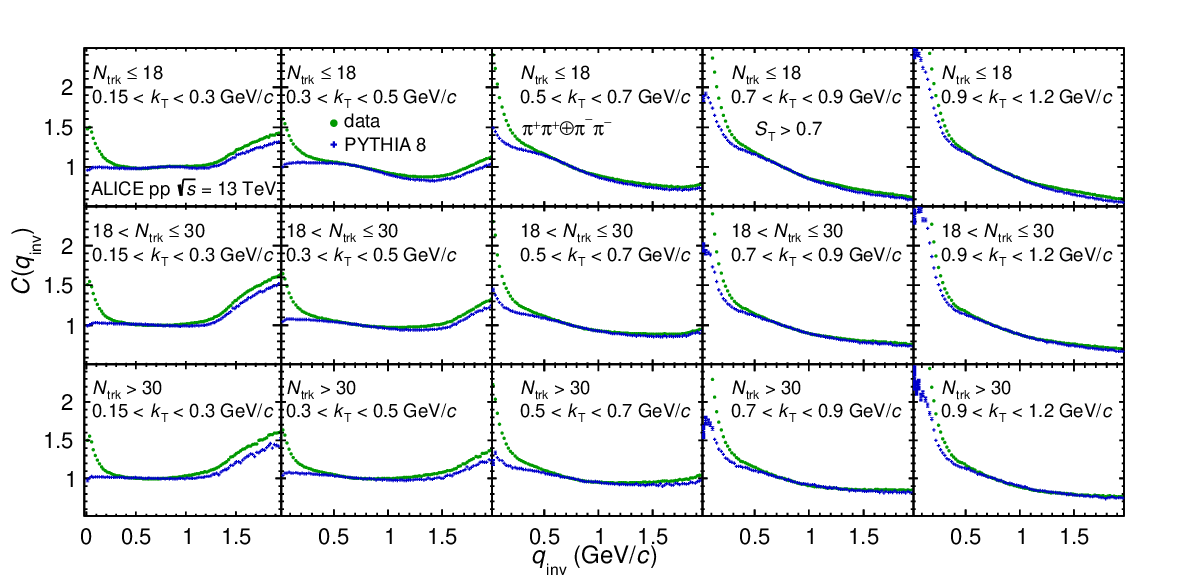}
\caption{The $\pi^\pm\pi^\pm$ experimental correlation functions (green solid circles) as function of the invariant pair relative momentum $q_\mathrm{inv}$ in pp collisions at $\sqrt{s}=13$~TeV for the raw multiplicity $N_{\rm trk}$ intervals of (1--18), (19--30), and ($>30$) in the (0.15--0.3)~GeV/$c$, (0.3--0.5)~GeV/$c$, (0.5--0.7)~GeV$/c$, (0.7--0.9)~GeV$/c$ and (0.9--1.2)~GeV$/c$ $k_{\rm T}$ intervals. A sphericity selection of $S_{\rm T} < 0.3$ is applied. The data are compared with PYTHIA 8 calculations shown by blue crosses. The error bars represent the statistical uncertainties, while the systematic uncertainties are negligible.}
\label{fig:CFsSp03}
\end{center}
\end{figure}

Figure~\ref{fig:CFsSp03} illustrates the pion experimental correlation function in pp collisions at $\sqrt{s}=13$~TeV compared with PYTHIA~8 model calculations for events with $S_{\rm T} < 0.3$. The CFs for jet-like events shown in this figure exhibit a pronounced slope over the full $q_\mathrm{inv}$ range, indicating the presence of non-femtoscopic effects. These effects are especially pronounced for $k_{\rm T} > 0.5$~GeV$/c$. As can be seen from the figure, the PYTHIA 8 model calculations describe reasonably well the pion experimental data for jet-like events at large $q_\mathrm{inv}$ values.

\subsubsection{Kaon correlation functions}\label{CF_KK}
\begin{figure}[h]
\begin{center}
\includegraphics[width=0.9\textwidth]{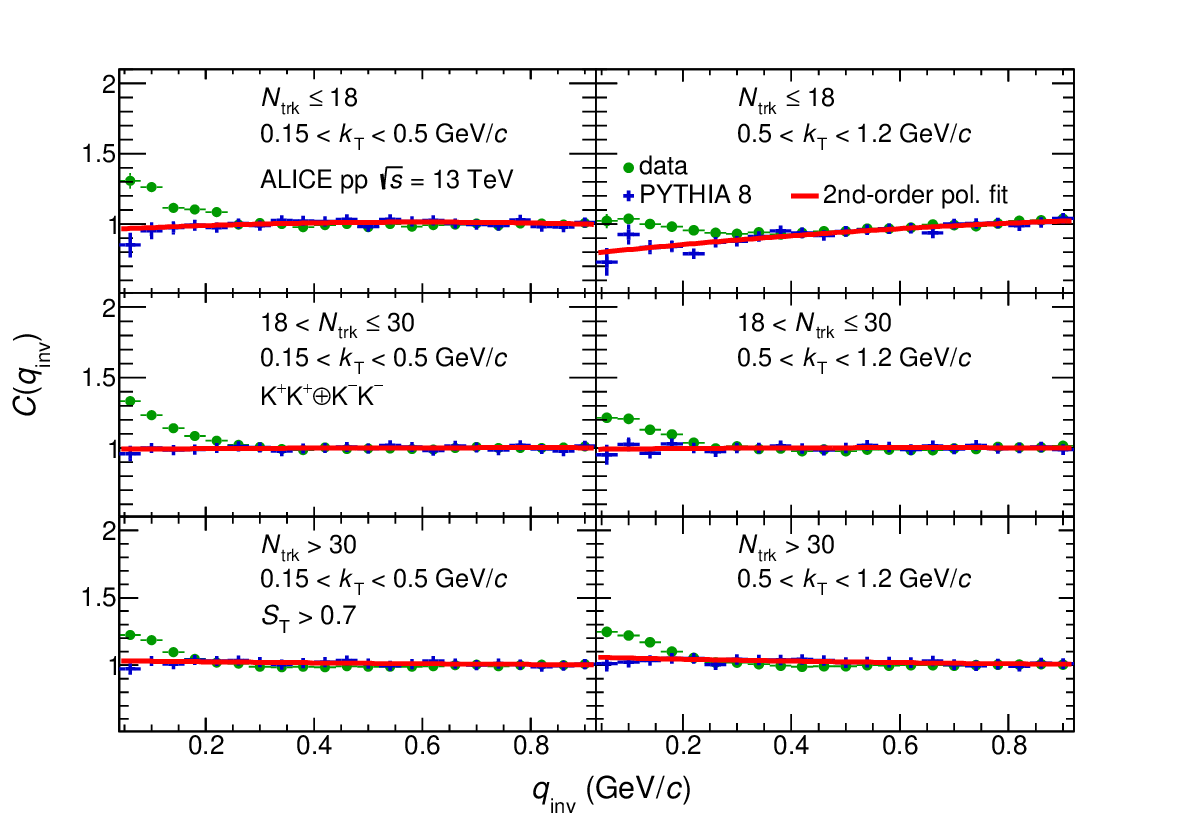}
\caption{The K$^{\pm}$K$^{\pm}$ experimental correlation functions (green solid circles) as function of the invariant pair relative momentum $q_\mathrm{\rm inv}$ in pp collisions at $\sqrt{s}=13$~TeV for the raw multiplicity $N_{\rm trk}$ intervals of (1--18), (19--30), and ($>30$) in the (0.15--0.5)~GeV$/c$ and (0.5--1.2)~GeV$/c$ $k_{\rm T}$ intervals. A sphericity selection of $S_{\rm T} > 0.7$ is applied. The data are compared with PYTHIA 8 calculations shown by blue crosses. The PYTHIA 8 calculations are approximated with a second-order polynomial (red curves). The error bars represent the statistical uncertainties, while the systematic uncertainties are negligible.
}
\label{fig:CFsKKSp07}
\end{center}
\end{figure}

Figure~\ref{fig:CFsKKSp07} shows the kaon experimental CF (green solid circles) for events with $S_{\rm T} > 0.7$ and the corresponding PYTHIA 8 calculations (blue crosses) in pp collisions at $\sqrt{s}=13$~TeV.

The strength of the charged kaon correlations, represented by the magnitude of $C(q_{\rm inv})$ for \mbox{$q_{\rm inv}\to 0$}, is smaller than observed for the pions and decrease with $k_{\rm T}$. The CFs for spherical events are flat at \mbox{$q_{\rm inv}>0.5$}~GeV$/c$ both for the data and the MC calculations. The non-femtoscopic background contributions obtained using PYTHIA 8 were fitted with a second-order polynomial, which then was used for the correction of the experimental CFs for non-femtoscopic effects. The fit allows reducing the impact of statistical fluctuations on the extracted femtoscopic parameters.

\begin{figure}[h]
\begin{center}
\includegraphics[width=0.9\textwidth]{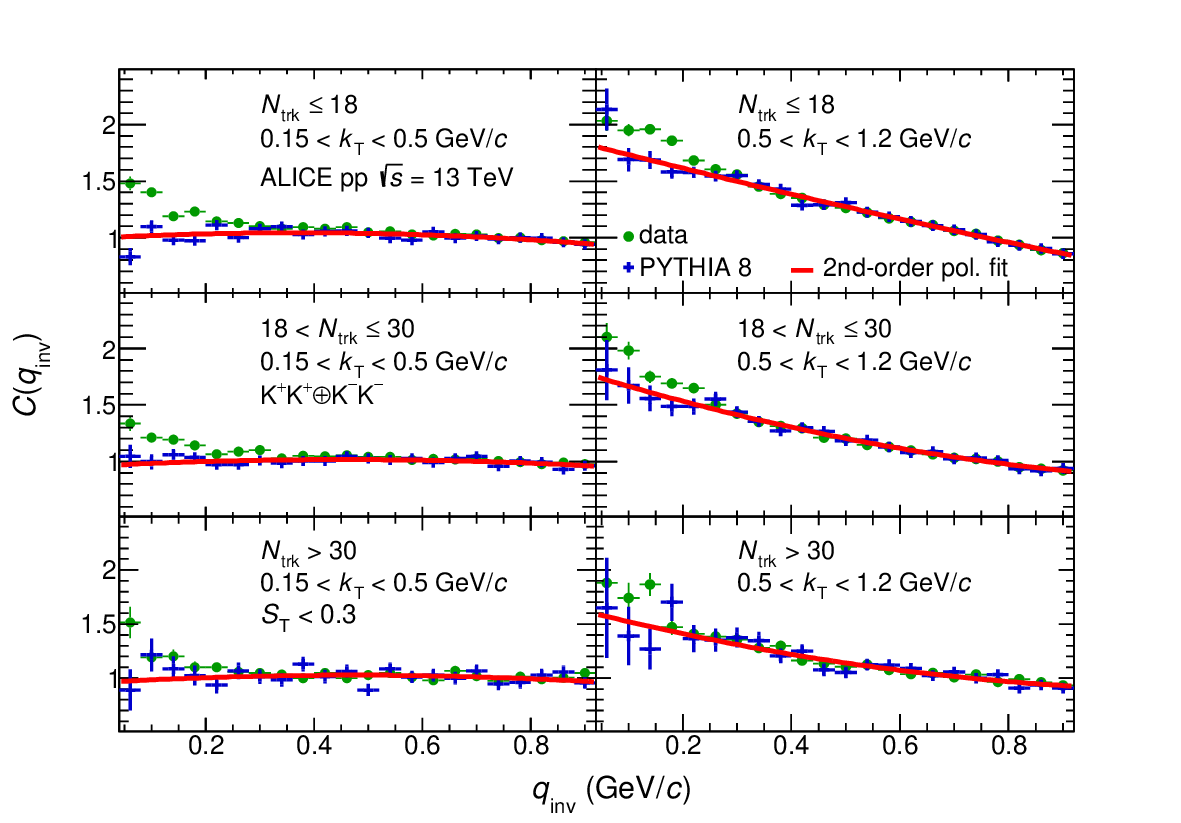}
\caption{
The K$^{\pm}$K$^{\pm}$ experimental correlation functions (green solid circles) as function of the invariant pair relative momentum $q_\mathrm{\rm inv}$ in pp collisions at $\sqrt{s}=13$~TeV for the raw multiplicity $N_{\rm trk}$ intervals of (1--18), (19--30), and ($>30$) in the (0.15--0.5)~GeV$/c$ and (0.5--1.2)~GeV$/c$ $k_{\rm T}$ intervals. A sphericity selection of $S_{\rm T} < 0.3$ is applied. The data are compared with PYTHIA 8 calculations, shown by blue crosses and approximated with a second-order polynomial (red curves). The error bars represent the statistical uncertainties, while the systematic uncertainties are negligible.
}
\label{fig:CFsKKSp03}
\end{center}
\end{figure}

Figure~\ref{fig:CFsKKSp03} presents the kaon experimental correlation function for events with $S_{\rm T} < 0.3$ in pp collisions at $\sqrt{s}=13$~TeV, for the raw multiplicity intervals $N_{\rm trk}$ of (1--18), (19--30), and ($>30$) in the (0.15--0.5)~GeV$/c$ and (0.5--1.2)~GeV$/c$ $k_{\rm T}$ intervals.
Similarly to the pion CFs for jet-like events (see Fig.~\ref{fig:CFsSp03}), the kaon jet-like CFs shown in Fig.~\ref{fig:CFsKKSp03} exhibit a pronounced slope at low $q_\mathrm{\rm inv}$, indicating the presence of non-femtoscopic effects. Such background is especially pronounced for \mbox{$k_{\rm T} > 0.5$}~GeV$/c$.
As can be seen in the figure, the estimate of the femtoscopic signal with respect to the background effects in the highest \mbox{$N_{\rm trk}>30$} multiplicity interval is not possible since the correlation functions coincide with PYTHIA 8 within statistical uncertainties.

\subsection{Correlation function parametrization}\label{FittingProcedure}
In the previous analyses performed by the ALICE Collaboration in pp collisions,
the Gaussian distribution of a particle source in the pair reference frame (PRF)
was assumed for pions~\cite{ALICE:2011kmy} and kaons~\cite{Abelev:2012sq}.
In those cases, the fit was performed using the Bowler--Sinyukov formula~\cite{Sinyukov:1998fc,Bowler:1986ta}
\begin{equation}
C(q_\mathrm{\rm inv})=N\left(1 -\lambda +\lambda K(r,q_\mathrm{\rm inv}) \times \left( 1+
\exp{\left(-R_\mathrm{\rm inv}^{2} q_\mathrm{\rm inv}^{2}\right)}\right)\right),
\label{eq:BS_CF}
\end{equation}
where $N$ is a normalization coefficient and $K(r,q_\mathrm{\rm inv})$ is the Coulomb function with a radius $r$ defined as
\begin{eqnarray}
K(r,q_\mathrm{\rm inv})=\frac{C({\rm QS+Coulomb})}{C({\rm QS})}.\label{Coulomb}
\end{eqnarray}
The parameters $R_\mathrm{\rm inv}$ and $\lambda$ describe the size of the
source and the correlation strength, respectively (see Eq.~(\ref{eq:BS_CF})).
The term $C({\rm QS})$ in Eq.~(\ref{Coulomb}) is a theoretical CF calculated with pure QS weights (wave function squared) and $C({\rm QS+Coulomb})$ corresponds to ${\rm QS+Coulomb}$ weights.
However, since the pion CFs are strongly non-Gaussian due to the large resonance contribution, an exponential Bowler--Sinyukov function was used to fit the pion CF, as in Ref.~\cite{ALICE:2019bdw}:
\begin{equation}
C(q_\mathrm{\rm inv})=N\left(1 -\lambda +\lambda K(r,q_\mathrm{\rm inv}) \times \left( 1+
\exp{\left(-R_\mathrm{\rm inv} q_\mathrm{\rm inv}\right)}\right)\right)D(q_{\rm inv}),
\label{eq:BSE_CF}
\end{equation}
where $D(q_{\rm inv})=bq_{\rm inv} + 1$ accounts for the slope of the baseline which remains after the division by PYTHIA 8.

\begin{figure}[h]
\begin{minipage}[h]{0.49\linewidth}
\center{\includegraphics[width=0.95\linewidth]{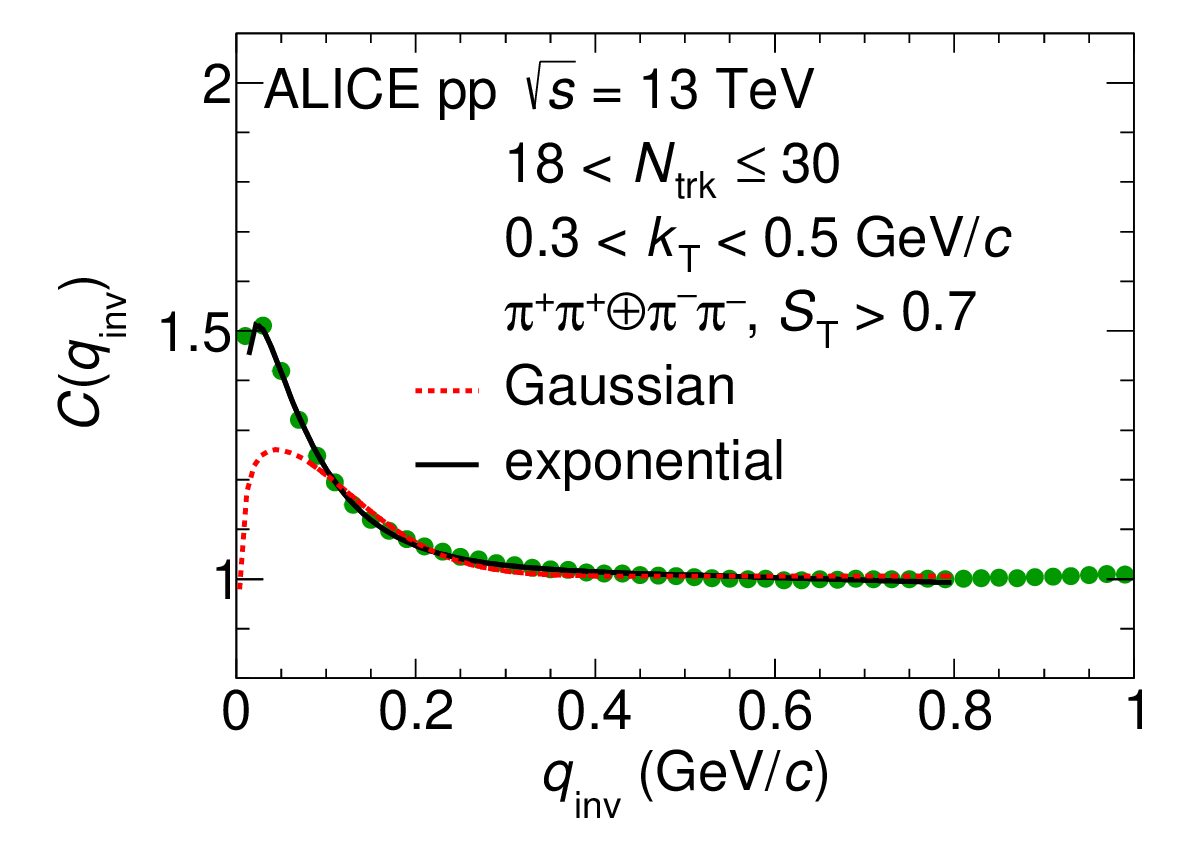}}
\end{minipage}
\hfill
\begin{minipage}[h]{0.49\linewidth}
\center{\includegraphics[width=0.95\linewidth]{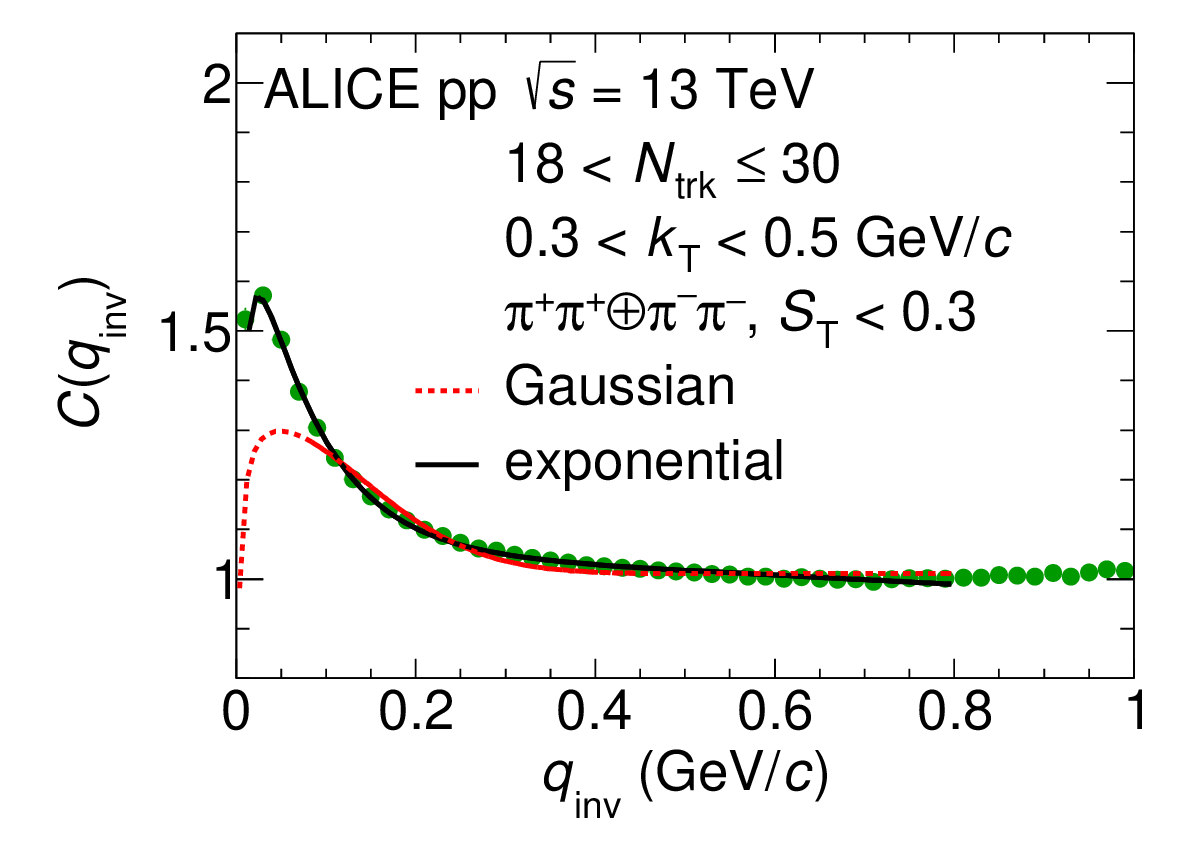}}
\end{minipage}
\caption{Examples of the pion CFs fitted with the Gaussian Eq.~(\ref{eq:BS_CF}) (dotted line) and the exponential (solid line) Bowler--Sinyukov Eq.~(\ref{eq:BSE_CF}) formulas for spherical (left panel) and jet-like (right panel) events.
Only statistical uncertainties are shown. Systematic uncertainties are negligible.
} \label{fig:CFsPiPiFit}
\end{figure}

The measured pion CFs shown in Figs.~\ref{fig:CFsSp07} and~\ref{fig:CFsSp03} were divided by the PYTHIA 8 baseline and fit with the exponential Bowler--Sinyukov formula of Eq.~(\ref{eq:BSE_CF}). Figure~\ref{fig:CFsPiPiFit} presents some examples of the pion CF fit with the Gaussian Bowler--Sinyukov formula of Eq.~(\ref{eq:BS_CF}) (dotted line) and the exponential one of Eq.~(\ref{eq:BSE_CF}) (solid line). The exponential fit function describes the pion CF well for both spherical and jet-like events, although the description is not ideal for \mbox{$q_{\rm inv} <0.05$}~GeV$/c$.

\begin{figure}[h]
\begin{minipage}[h]{0.49\linewidth}
\center{\includegraphics[width=0.95\linewidth]{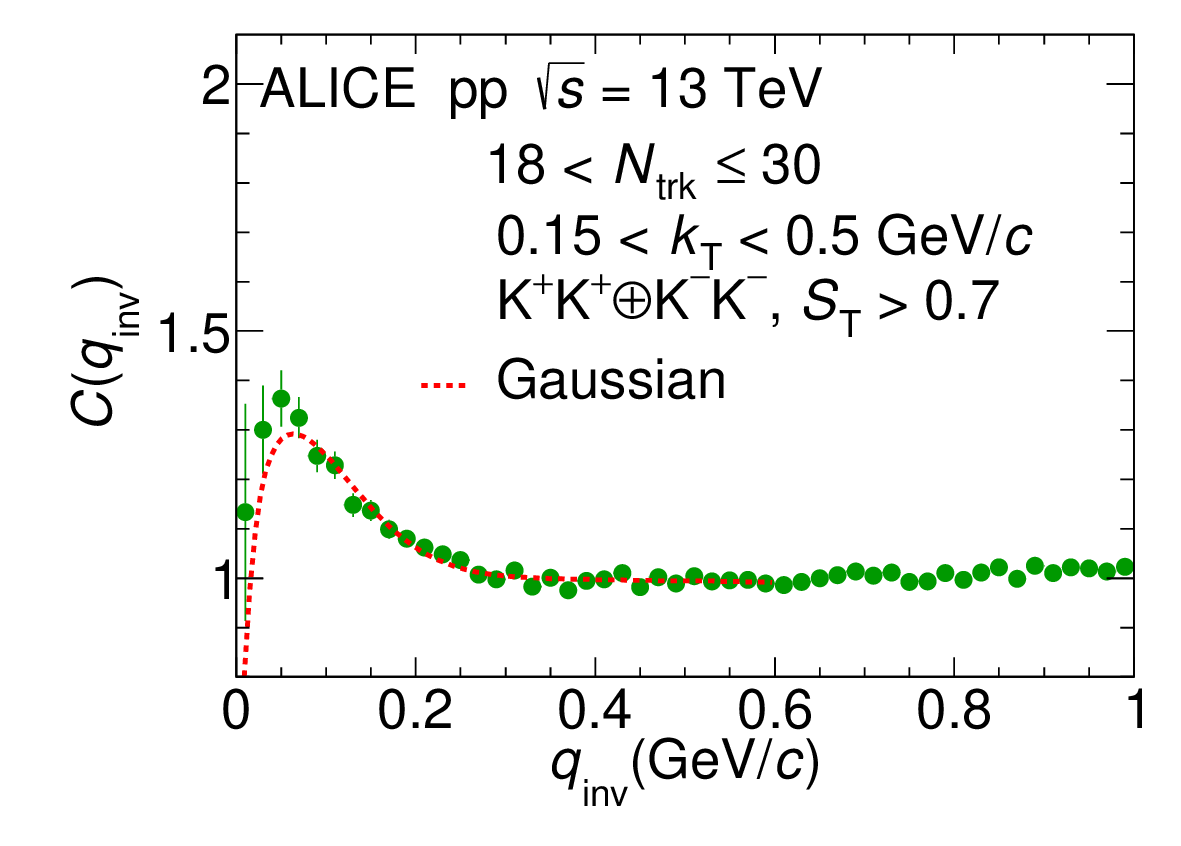}}
\end{minipage}
\hfill
\begin{minipage}[h]{0.49\linewidth}
\center{\includegraphics[width=0.95\linewidth]{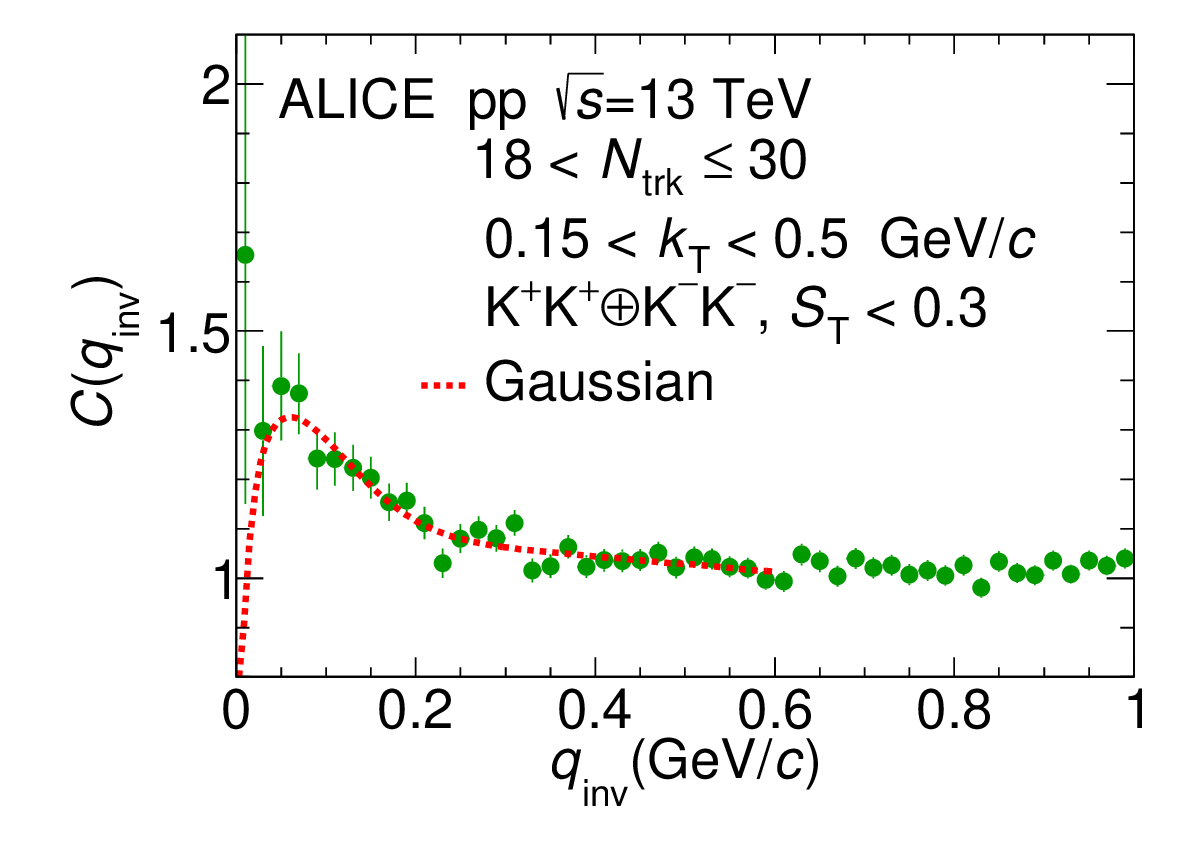}}
\end{minipage}
\caption{Examples of the kaon CFs fitted with the Gaussian Eq.~(\ref{eq:BS_CF}) (dotted line) Bowler--Sinyukov formula for spherical (left panel) and jet-like (right panel) events.
Only statistical uncertainties are shown. Systematic uncertainties are negligible.
} \label{fig:CFsKKFit}
\end{figure}

The kaon CFs for spherical and jet-like events were corrected using a second-order polynomial to describe the non-femtoscopic background as explained above and fitted with the Gaussian Bowler--Sinyukov formula (Eq.~(\ref{eq:BS_CF})). An example of such a fit is shown in Fig.~\ref{fig:CFsKKFit} for both spherical and jet-like events.
For the largest multiplicity bin for $S_{\rm T} < 0.3$ selection, kaon CFs do not exhibit any femtoscopic peak after the correction for the baseline and, therefore, the fit was not performed.


\subsection{Systematic uncertainties}\label{SystematicUncertainties}
In this section, the systematic uncertainties estimated for the extracted source parameters $R_{\rm inv}$ and $\lambda$ for $\pi^\pm\pi^\pm$ and K$^{\pm}$K$^{\pm}$ correlations at different sphericities are discussed.

The systematic uncertainty contributions related to particle selection and fit criteria \mbox{$\Delta_{\rm sys} = |y_{0} - y_{\rm var}|$} were considered, taking into account their statistical significance level determined by the Barlow factor~\cite{Barlow:2002yb}
\begin{equation}
B = \frac{|y_{0}-y_{\rm var}|}{\sqrt{\sigma_{0}^{2}+\sigma_{\rm var}^{2}-2\rho\sigma_{0}\sigma_{\rm var}}},
\label{eq:Barlow}
\end{equation}
where $y_{0}$ is the default value, $y_{\rm var}$ is a value obtained with some variation of either the selection criteria or fit conditions, $\sigma_{0}$ is the statistical uncertainty of the default value, $\sigma_{\rm var}$ is the statistical uncertainty of the femtoscopic parameters obtained using varied analysis criteria, and $\rho$ characterizes the correlation between $y_0$ and $y_{\rm var}$.

The systematic uncertainty contribution is taken into account if $B>1$, i.e.\ only systematic uncertainties whose statistical significance level exceeds 68\% are included in the total systematic uncertainty value. The systematic uncertainties $\Delta^i_{\rm sys}$, corresponding to each variation $i$ of the particle selection and fit criteria and having $B>1$, were added in quadrature to give the total systematic uncertainty $\Delta$ value
\begin{equation}
\Delta_{\rm sys} = \sqrt{\sum_i\left(\Delta_{\rm sys}^i\right)^{2}}.
\label{eq:toterr}
\end{equation}

As event selection criterion variation, the sphericity intervals were varied by $\pm0.05$~\cite{ALICE:2019bdw}, resulting in a systematic uncertainty of up to 10\% for the radii and $\lambda$ parameters for pions and kaons.

The systematic uncertainties related to the track selection ($p_{\rm T}$ and $\eta$ intervals, $\rm DCA_{\rm transverse}$, $\rm DCA_{\rm longitudinal}$, the values of the number of track points in TPC) were estimated by varying each selection criterion (see Tables~\ref{tab:PiPicuts} and~\ref{tab:KKcuts}) with variation limits up to $\pm20\%$. Another source of systematic uncertainty is the misidentification of particles and the associated purity correction. The same $\pm20\%$ variation of the parameters $n_{\sigma_{\mathrm{TPC}}}$ and $n_{\sigma_{\mathrm{TOF}}}$
used for the purity correction estimation was performed. The purity correction influences only the $\lambda$ parameter. For pions, single particle purity is $\approx$~99\% at $p < 1.5$ GeV$/c$, so the correction for purity is negligible. For kaons, it is $<$~1\%. The estimated systematic uncertainty for pions for spherical events is 2--5\% for $R_{\rm inv}$ and $\lambda$, while for jet-like events it is 1--15\% for $R_{\rm inv}$ and 1--20\% for $\lambda$. For kaons, it is $<5$\% for spherical events, and 10--20\% for jet-like events for both $R_{\rm inv}$ and $\lambda$.

To minimize the influence of two-track effects, the tracks in this analysis were required to have an average TPC separation of at least 3~cm. The systematic uncertainty related to these effects was estimated from comparing the resulting femtoscopic parameters with those obtained with the average TPC separation up to 10~cm. This comparison showed that the influence of the two-track effects on the extracted parameters was negligible.

It is known that usually the extracted femtoscopic parameters noticeably depend on the fit range if the baseline is not flat, which is the case for CF in pp collisions.
To estimate the related systematic uncertainty, three different fit ranges are considered: the standard \mbox{$0<q_{\rm inv}<0.8$}~GeV$/c$, the shorter \mbox{$0<q_{\rm inv}<0.7$}~GeV$/c$, and the longer \mbox{$0<q_{\rm inv}<0.9$}~GeV$/c$ ones. The systematic uncertainty for pion and kaon due to the variation in fit range is 1--12\% for $R_{\rm inv}$ and 5--20\% for $\lambda$.

The Coulomb interaction, described by the factor $K$ in Eq.~(\ref{Coulomb}), is included in the default fitting procedure (Eqs.~(\ref{eq:BS_CF})--(\ref{eq:BSE_CF})). The influence of strong interactions on the extracted femtoscopic observables is estimated using the factor $K(q_\mathrm{\rm inv})=C({\rm QS+Coulomb+strong})/C({\rm QS})$ and is taken as a systematic uncertainty.
The strong interaction for both pions and kaons was calculated using the fully dynamical lattice QCD result~\cite{Beane:2007uh}.
The resulting systematic uncertainty for pions is 3--7\% for $R_{\rm inv}$ and $\lambda$ for both spherical and jet-like events.
The systematic uncertainty related to the strong interactions for kaons is 7--15\% for radii and correlation strengths for both spherical and jet-like events.

The systematic uncertainty introduced by the fit range variation was the only one observed to be fully correlated. For this uncertainty $\rho=1$ in Eq.~(\ref{eq:Barlow}), while for all other uncertainties $\rho=0$ was used.

\section{Results and discussion}\label{ResultsDiscussion}

One of the goals of this work is to compare the kaon and pion radii for spherical and jet-like events in order to understand if the extracted radii follow the same $m_{\rm T}$ scaling behavior as observed in heavy-ion collisions.

To compare the strongly non-Gaussian pion source with the Gaussian kaon one, the exponential pion radii have to be converted to a Gaussian form. The Gaussian parametrization assumes a radial Gaussian distribution with the first moment of the distribution $1/R\sqrt{\pi}$, while the exponential parametrization assumes a Lorentzian one with the first moment of distribution $1/R$. Therefore, in order to compare the radii, the exponential radii should be divided by $\sqrt{\pi}$.

\begin{figure}[h]
\begin{center}
\includegraphics[width=0.99\textwidth]{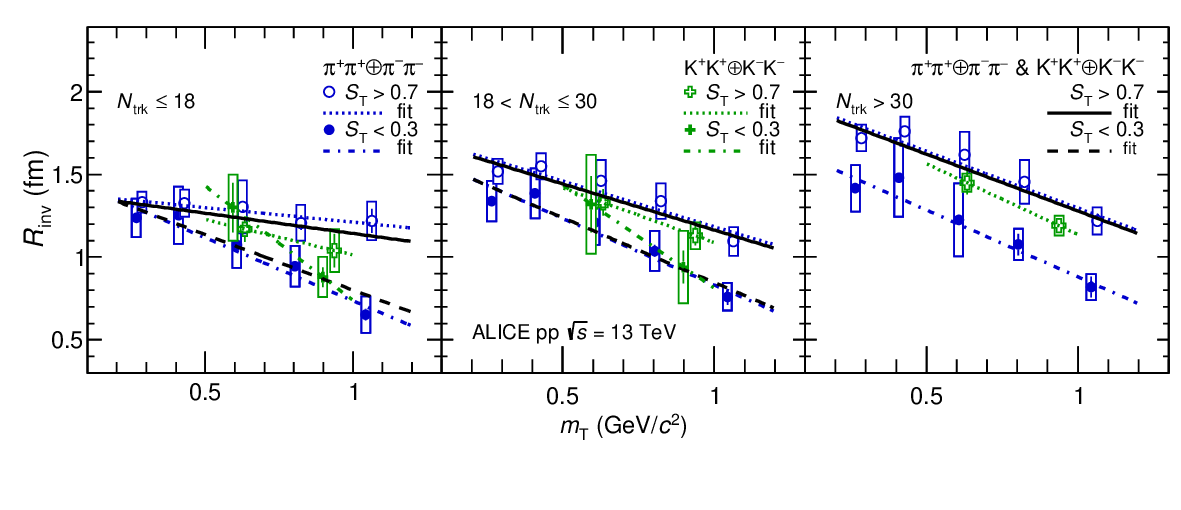}
\vspace*{-0.9cm}
\caption{The Gaussian pion (blue circles) and kaon (green crosses) radii for spherical ($S_{\rm T}>0.7$) and jet-like ($S_{\rm T}<0.3$) events as function of the average pair transverse mass $m_{\rm T}$ for different multiplicity intervals. The lines approximating the pion and kaon radii by the linear function of Eq.~(\ref{eq:linfit}) are shown for spherical events by dotted blue and green lines, respectively, and for jet-like events by dotted-dashed lines. The lines for combined fit for pion and kaon points are shown by black solid and dashed lines for spherical and jet-like events, respectively. Statistical (bars) and systematic (boxes) uncertainties are shown.}
\label{fig:RmtSp07Sp03}
\end{center}
\end{figure}
\begin{figure}[h]
\begin{center}
\includegraphics[width=0.99\textwidth]{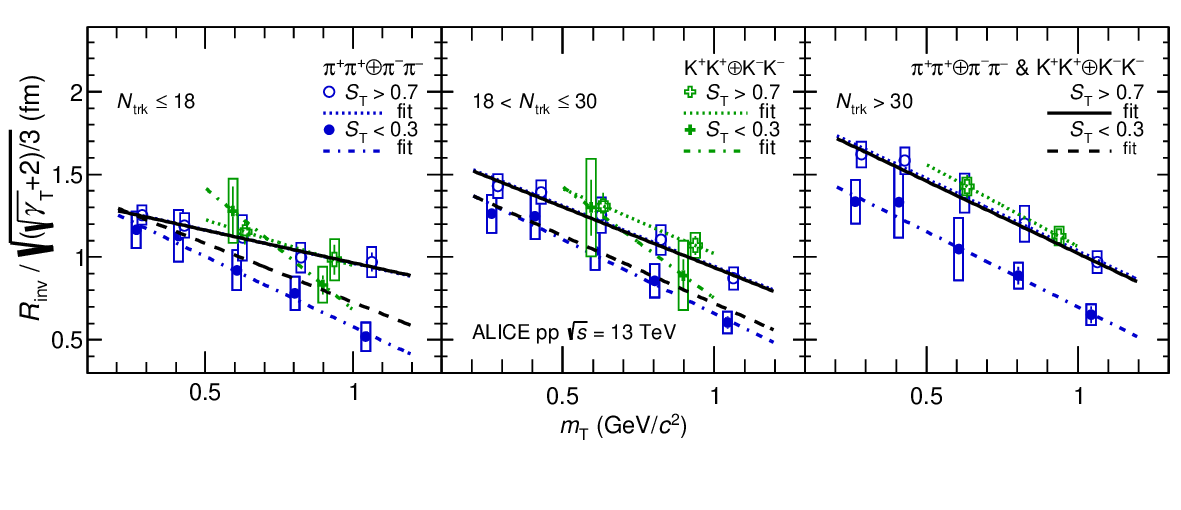}
\vspace*{-0.9cm}
\caption{The Gaussian pion (blue circles) and kaon (green crosses) radii for spherical ($S_{\rm T}>0.7$) and jet-like ($S_{\rm T}<0.3$) events corrected for the Lorenz boost with factor in Eq.~(\ref{eq:corr_fact}) as function of the average pair transverse mass $m_{\rm T}$ for different multiplicity intervals. The lines approximating the pion and kaon radii by the linear function of Eq.~(\ref{eq:linfit}) are shown for spherical events by dotted blue and green lines, respectively, and for jet-like events by dotted-dashed lines. The lines for combined fit for pion and kaon points are shown by black solid and dashed lines for spherical and jet-like events, respectively. Statistical (bars) and systematic (boxes) uncertainties are shown.}
\label{fig:corrRmtSp07Sp03}
\end{center}
\end{figure}
Figure~\ref{fig:RmtSp07Sp03} presents the $m_{\rm T}$ dependence of Gaussian radii for spherical and jet-like events for pions and kaons in different raw multiplicity intervals. As discussed in Section~\ref{CF_pi}, the pion radii for \mbox{$S_{\rm T}>0.7$}, \mbox{$N_{\rm trk}<$}18, and \mbox{$k_{\rm T}>0.50$}~GeV$/c$ (\mbox{$m_{\rm T}>~0.52$}~GeV$/c^2$) have to be interpreted with caution because the baseline used to describe the non-femtoscopic effects was distorted at low $q_{\rm inv}$ (see Fig.~\ref{fig:CFsSp07}) by the \mbox{$p_{\rm T}>0.5$}~GeV$/c$ three-track requirement requested for the calculation of the transverse sphericity.

The pion and kaon radii for spherical and jet-like events (Figures~\ref{fig:RmtSp07Sp03} and~\ref{fig:corrRmtSp07Sp03}) were fitted in each multiplicity interval with a linear function,
\begin{equation}
f(m_{\rm T}) = a + bm_{\rm T},
\label{eq:linfit}
\end{equation}
where $a$ and $b$ are free parameters. The results of this fitting are presented in Tables~\ref{tab:SlopesNotCorr} and~\ref{tab:SlopesCorr}.
The individual fits for pions and kaons are compared with the combined fit of pion and kaon radii. It should be noted that for kaons the line is not exactly a fit since there are only two points available, but it illustrates the slope of the radii with $m_{\rm T}$. It allows quantifying how much the kaon radius data points deviate from this fit and, thus, how well the approximate \mbox{$m_{\rm T}$} scaling manifests itself. Both the spherical ($S_{\rm T}>0.7$) and the jet-like ($S_{\rm T}<0.3$) radii for pions and kaons decrease with increasing $m_{\rm T}$. The statistical uncertainties are large, especially for kaons, but there is an indication that the spherical radii demonstrate a flatter dependence compared to the jet-like ones for $N_{\rm trk}<30$, while a more pronounced slope appears for $N_{\rm trk}>30$ (see Table~\ref{tab:SlopesNotCorr}). The spherical kaon radii are smaller than the corresponding spherical pion radii, and the difference increases with increasing multiplicity. The jet-like pion radii are smaller than spherical ones, and the difference increases with increasing multiplicity.

\begin{table}
\centering
\caption{The parameters of the approximation by the linear function of Eq.~\ref{eq:linfit} of the femtoscopic radii in PRF calculated for pions and kaons for spherical and jet-like events, as a function of pair transverse mass for the different multiplicity intervals from Fig.~\ref{fig:RmtSp07Sp03}.}
\begin{tabular}{lcccc}
\hline
\hline
 type of fit & $N_{\rm trk}$ &  $b$  & $a$ &  $\chi^{2}/{\rm NDF}^{{\phantom{1}}^{\phantom{1}}}$     \\
\hline
 & 3(1)--18&     $-0.18\pm$ 0.16   &  1.39  $\pm$ 0.09   &  0.18/3   \\
pions, spherical events & 19--30  &     $-0.55\pm$ 0.14   &  1.74  $\pm$ 0.09   &  1.95/3   \\
 & $>30$    &     $-0.69\pm$ 0.14   &  1.98  $\pm$ 0.10   &  1.77/3  \\
\hline
 & 3(1)--18&     $-0.76\pm$ 0.20   &  1.50  $\pm$ 0.14   &  0.90/3   \\
pions, jet-like events & 19--30  &      $-0.80\pm$ 0.14   &  1.64  $\pm$ 0.14   &  1.2/3 \\
 & $>30$    &      $-0.81\pm$ 0.21   &  1.69  $\pm$ 0.17   &  0.63/3   \\
\hline
\hline
 & 3(1)--18&    $-0.43\pm$ 0.60   &  1.44  $\pm$ 0.42    &  -   \\
kaons, spherical events & 19--30  &     $-0.66\pm$ 0.43  &  1.75  $\pm$ 0.34    &  -   \\
 & $>30$    &     $-0.85\pm$ 0.38  &  1.99  $\pm$ 0.31    &  -    \\
\hline
  & 3(1)--18&    $-1.38\pm$ 0.93   &  2.12  $\pm$ 0.78    &  -   \\
kaons, jet-like events  & 19--30  &    $-1.25\pm$ 1.38   &  2.06  $\pm$ 1.12    &  -   \\
 & $>30$    &   - &  -  &  -   \\
\hline
\hline
 & 3(1)--18&    $-0.24\pm$ 0.15   &  1.39 $\pm$ 0.09    &  2.10/5  \\
pions and kaons, spherical events & 19--30  &    $-0.56\pm$ 0.14   &  1.72  $\pm$ 0.09   &  2.25/5   \\
 & $>30$    &     $-0.69\pm$ 0.14  &  1.97  $\pm$ 0.09   &  2.00/5 \\
\hline
 & 3(1)--18&    $-0.67\pm$ 0.16   &  1.47  $\pm$ 0.14    &  1.60/5   \\
pions and kaons, jet-like events & 19--30  &     $-0.78\pm$ 0.17  &  1.63  $\pm$ 0.14   &  1.30/5   \\
 & $>30$    &   - &  -  &  -   \\
\hline\hline
\end{tabular}
\label{tab:SlopesNotCorr}
\end{table}

The $m_{\rm T}$ scaling for pions and kaons was predicted in a case of negligible transverse flow and common freeze-out for 3D radii in the Longitudinally Co-Moving System (LCMS)~\cite{Makhlin:1987gm}), where the pair relative momentum is decomposed over ($q_{\rm out}$, $q_{\rm side}$, $q_{\rm long}$). Here, the ``long'' component goes along the beam direction, ``out'' goes along the pair transverse momentum, and ``side'' goes perpendicular to the latter in the transverse plane, while the longitudinal total pair momentum vanishes. Theoretical calculations within the 3+1D hydrodynamic model coupled with the statistical hadronization code THERMINATOR-2 taking into account the resonance contribution showed an effective power-law scaling of 3D LCMS radii over the pair transverse mass for pions, kaons, and protons~\cite{Kisiel:2014upa}. The same scaling was also observed in Ref.~\cite{Kisiel:2014upa} for the one-dimensional radii in the RPF. However, it is often challenging to do measurements in the LCMS due to limited number of events and, therefore, the measurements are performed in the PRF and then transformed from the PRF to the LCMS by applying a Lorentz boost in the direction of the pair transverse momentum with velocity \mbox{$\beta_{\rm T} = p_{\rm T}/m_{\rm T}$} as $\gamma_{\rm T} R_{\rm out}$ (where \mbox{$\gamma_{\rm T}=\sqrt{1-\beta_{T}^2}$)}. Therefore, the transverse (out) component of the 3D radius changes differently for pions and for kaons due to the different Lorentz boosts. In Ref.~\cite{Kisiel:2014upa}, it was shown that the scaling could be restored if the radii were divided by the following scaling factor
\begin{equation}
f=\sqrt{(\sqrt\gamma_{\rm T}+2)/3}.
\label{eq:corr_fact}
\end{equation}
The factor $f$ was estimated through numerical simulations reproducing the 3D radius growth with $\gamma_{\rm T}$.
After applying these kinematic corrections (Eq.~(\ref{eq:corr_fact})), the authors of Ref.~\cite{Kisiel:2014upa} observed that the one-dimensional  $R_{\rm inv}$ correlation radii for pions, kaons, and protons, as measured by ALICE in Pb--Pb collisions at  {\ensuremath{\sqrt{s_{\rm NN}}}{~=~2.76}~TeV}~\cite{Adam:2015vja}, lied on a common curve (with the accuracy of 10\%). The effect of the factor $f$ was found to be the same for CFs with different non-Gaussian tails since the extracted radii values are mainly determined by the fit in the region of the femtoscopic peak. Therefore, this effect could be similar in collisions of different types.

The simplest way to see the possible difference in the pp collisions is to apply  this correction to the radii measured in this work. Figure~\ref{fig:corrRmtSp07Sp03} shows that the extracted pion and kaon radii for spherical events become closer to each other than in the case without such correction (Fig.~\ref{fig:RmtSp07Sp03}) considered above.
The obtained $m_{\rm T}$ scaling of the pion and kaon radii is not as good as for the Pb--Pb case~\cite{Kisiel:2014upa}, which can be explained by different influence of resonance decay contributions in pp and Pb--Pb collisions.
The large uncertainties for kaon radii do not allow making any conclusion about the $m_{\rm T}$ scaling in jet-like events. The understanding of the $m_{\rm T}$ trend of the jet-like pion and kaon radii requires further study. The radii shown in Fig.~\ref{fig:corrRmtSp07Sp03} were also fitted with the linear function of Eq.~\ref{eq:linfit} (see Table~\ref{tab:SlopesCorr}). The $\chi^{2}/{\rm NDF}$ for the fit of the uncorrected radii (Table~\ref{tab:SlopesNotCorr}) are larger than for the corrected ones (Table~\ref{tab:SlopesCorr}) for spherical events in all multiplicity intervals.
\begin{table}
\centering
\caption{The parameters of the approximation by the linear function of Eq.~\ref{eq:linfit} of the femtoscopic radii in the PRF calculated for pions and kaons for spherical and jet-like events, as a function of pair transverse mass for the different multiplicity intervals from Fig.~\ref{fig:corrRmtSp07Sp03}.}
\begin{tabular}{lcccc}
\hline\hline
 Type of fit & $N_{\rm trk}$ &  $b$  & $a$ &  $\chi^{2}/{\rm NDF}^{{\phantom{1}}^{\phantom{1}}}$     \\
\hline
 & 3(1)--18&     $-0.40\pm$ 0.13   &  1.36  $\pm$ 0.08   &  0.24/3   \\
pions, spherical events & 19--30  &     $-0.73\pm$ 0.12   &  1.68  $\pm$ 0.08   &  0.81/3   \\
 & $>30$    &     $-0.87\pm$ 0.12   &  1.91  $\pm$ 0.09   &  0.69/3  \\
\hline
 & 3(1)--18&     $-0.85\pm$ 0.17   &  1.43  $\pm$ 0.13   &  0.39/3   \\
pions, jet-like events & 19--30  &      $-0.90\pm$ 0.16   &  1.55  $\pm$ 0.12   &  0.51/3 \\
 & $>30$    &      $-0.91\pm$ 0.19   &  1.61  $\pm$ 0.15   &  0.27/3   \\
\hline
\hline
 & 3(1)--18&    $-0.55\pm$ 0.56   &  1.50  $\pm$ 0.40    &  -   \\
kaons, spherical events & 19--30  &     $-0.79\pm$ 0.42  &  1.81  $\pm$ 0.33    &  -   \\
 & $>30$    &     $-0.99\pm$ 0.37  &  2.05  $\pm$ 0.31    &  -    \\
\hline
  & 3(1)--18&    $-1.47\pm$ 0.9   &  2.15  $\pm$ 0.76    &  -   \\
kaons, jet-like events  & 19--30  &    $-1.35\pm$ 1.35   &  2.10  $\pm$ 0.10    &  -   \\
 & $>30$    &   - &  -  &  -   \\
\hline
\hline
 & 3(1)--18&    $-0.39\pm$ 0.13   &  1.36 $\pm$ 0.08    &  0.45/5  \\
pions and kaons, spherical events & 19--30  &    $-0.70\pm$ 0.12   &  1.67  $\pm$ 0.08   &  1.25/5   \\
 & $>30$    &     $-0.87\pm$ 0.12  &  1.90  $\pm$ 0.09   &  0.75/5 \\
\hline
 & 3(1)--18&    $-0.71\pm$ 0.14   &  1.44  $\pm$ 0.12    &  10.1/5   \\
pions and kaons, jet-like events & 19--30  &     $-0.82\pm$ 0.15  &  1.54  $\pm$ 0.12   &  3.10/5   \\
 & $>30$    &   - &  -  &  -   \\
\hline\hline
\end{tabular}
\label{tab:SlopesCorr}
\end{table}
This suggests an approximate $m_{\rm T}$ scaling for the radii for spherical events if the differences in Lorentz boosts for pions and kaons are taken into account.

\begin{figure}[h]
\begin{center}
\includegraphics[width=0.95\textwidth]{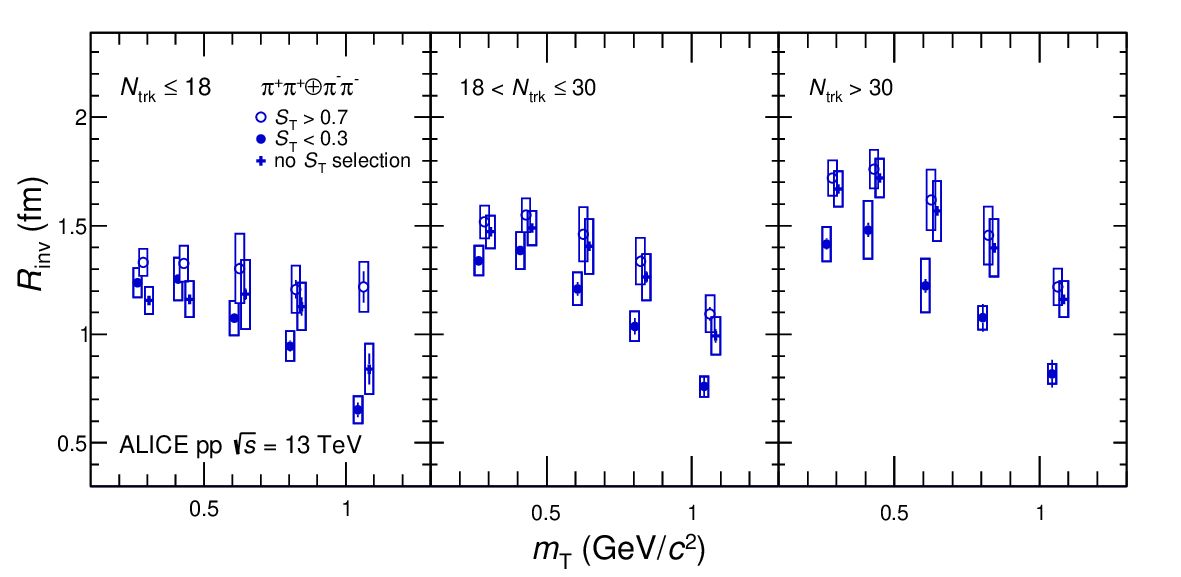}
\caption{The Gaussian pion radii with $S_{\rm T}>0.7$ (open circles), $S_{\rm T}<0.3$ (solid circles) and without sphericity selection (solid crosses). Statistical (bars) and systematic (boxes) uncertainties are shown.}
\label{fig:RmtSp07Sp03ALL}
\end{center}
\end{figure}

It is also instructive to consider the difference between the radii with and without sphericity selection, as shown in Fig.~\ref{fig:RmtSp07Sp03ALL} for the pion radii for $S_{\rm T}>0.7$, $S_{\rm T}<0.3$, and without a selection on sphericity.
It can be seen that for large multiplicities ($N_{\rm trk}>18$) the radii extracted from the data without sphericity selections are close to those obtained in spherical events. This can be naturally explained by the experimental sphericity distributions presented in Fig.~\ref{fig:Sphericity}. The figure shows that all events tend to have large value of $S_{\rm T}$ and that the high-multiplicity intervals consist mainly of events with $S_{\rm T}>0.7$. Therefore, the corresponding radii are close to each other within uncertainties. For the $N_{\rm trk}<18$ interval, the fraction of particles with $S_{\rm T}<0.3$ becomes large, and the radii calculated without sphericity selection for this multiplicity interval are in between the $S_{\rm T}>0.7$ and $S_{\rm T}<0.3$ sphericity selected results, reflecting some interplay between all sphericity selected contributions.

\begin{figure}[h]
\begin{minipage}[h]{0.49\linewidth}
\center{\includegraphics[width=0.95\linewidth]{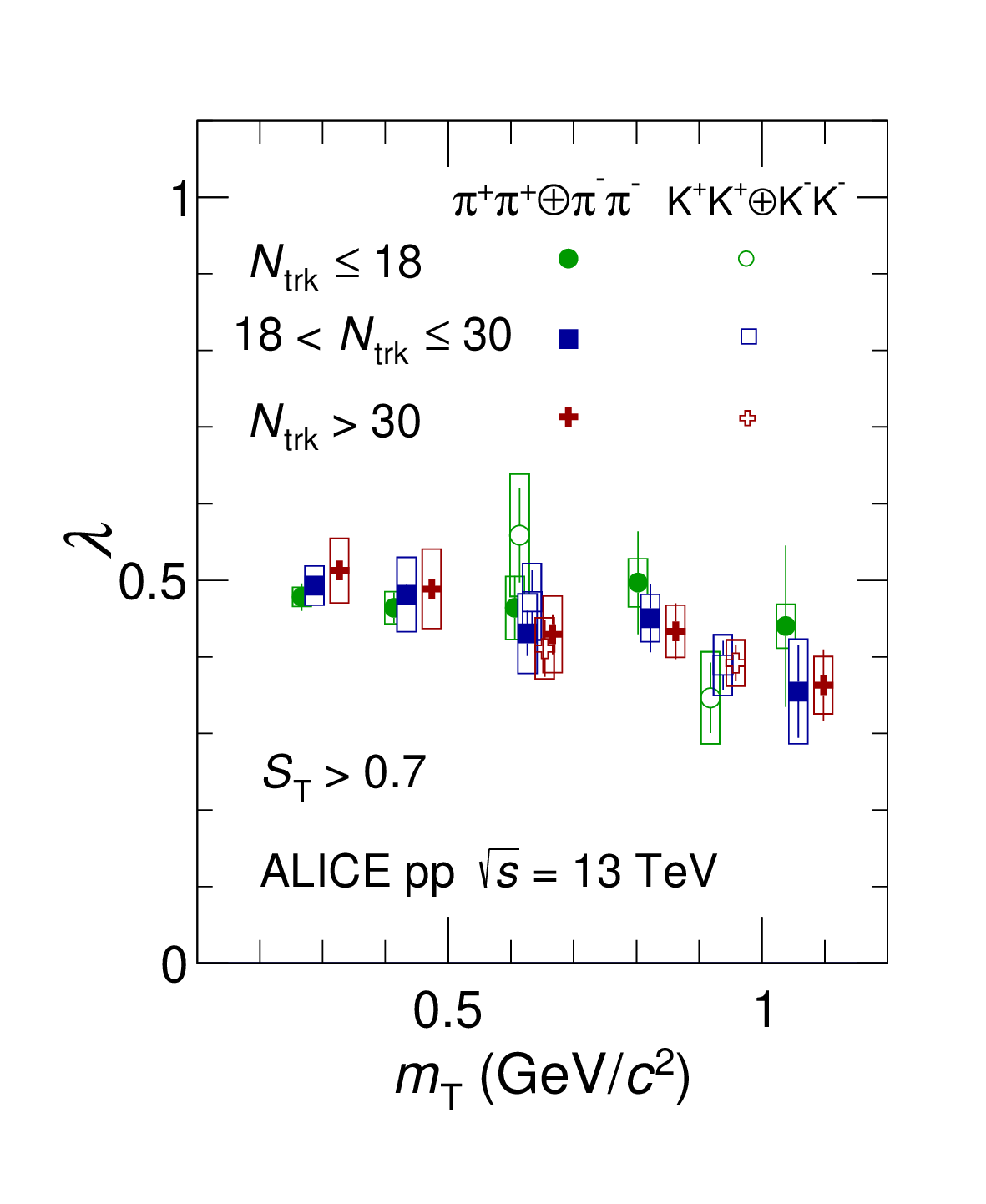}}
\end{minipage}
\hfill
\begin{minipage}[h]{0.49\linewidth}
\center{\includegraphics[width=0.95\linewidth]{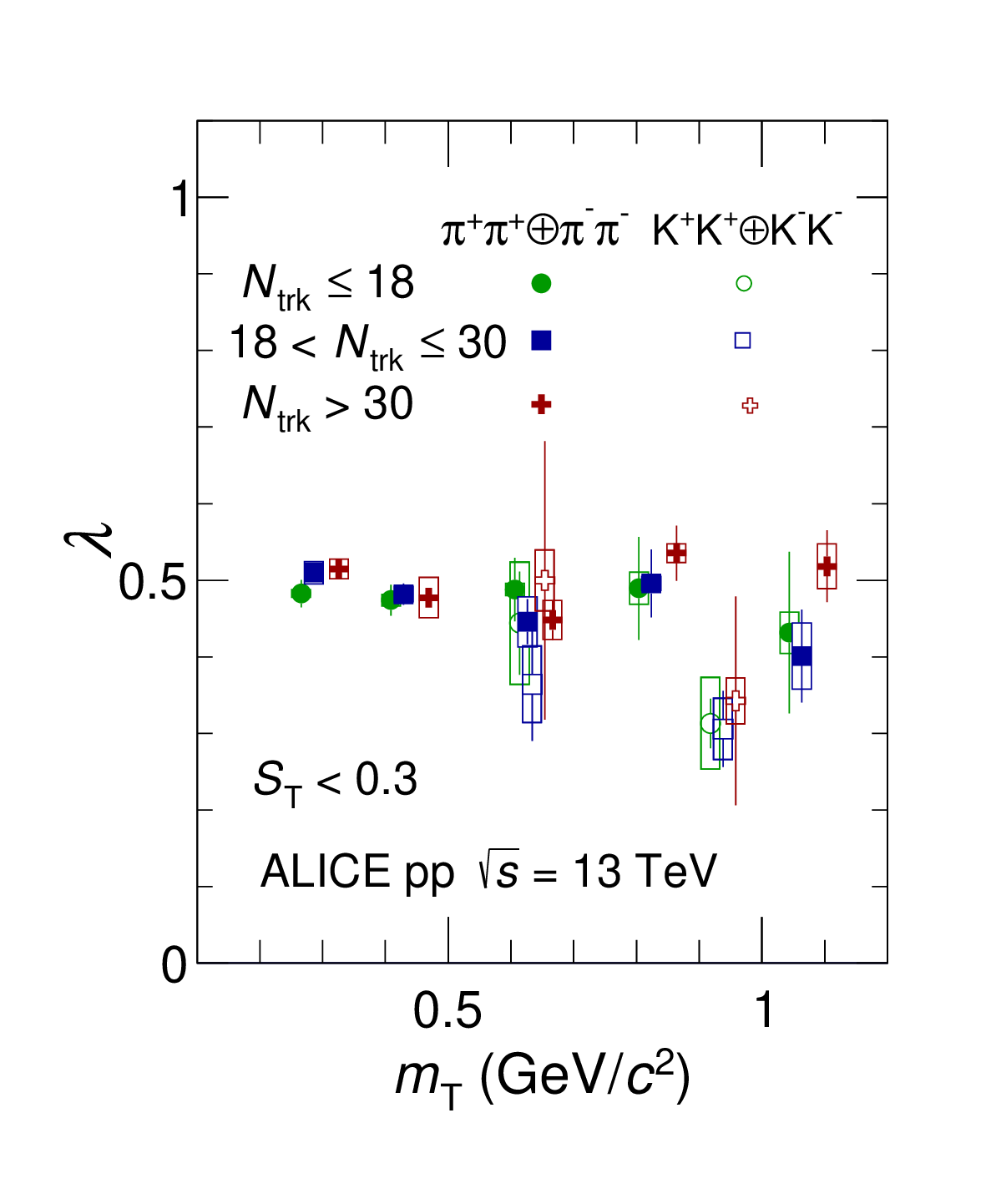}}
\end{minipage}
\caption{The Gaussian pion and kaon $\lambda$ parameters for spherical (left panel) and jet-like (right panel) events as a function of the average pair transverse mass $m_{\rm T}$ for different multiplicity intervals. The $\lambda$ parameters are corrected for purity. Statistical (bars) and systematic (boxes) uncertainties are shown.
} \label{fig:Lambda}
\end{figure}

Figure~\ref{fig:Lambda} demonstrates the $m_{\rm T}$ dependence of the correlation strength parameters $\lambda$ for pions and kaons for both spherical (left panel) and jet-like (right panel) events. The $\lambda$ parameters for pions were converted to a Gaussian form by dividing them by~$\sqrt{\pi}$~in the same way as was done for the radii (see the discussion in the beginning of this section). The obtained pion $\lambda$ weakly decreases with increasing $m_{\rm T}$, and this behavior is similar for both sphericity selections. The kaon $\lambda$ parameters are close to the pion ones, and all of them are in the range of $\approx$0.4--0.5. The main factors that could decrease the $\lambda$ parameters compared to the ideal case of unity are the non-Gaussian shape of CFs due to contribution of particles from short-lived (strongly decaying) and medium-lived ($\omega$ for pions, ${\rm K}^{0}_{\rm S}$ for kaons) resonances and the $R_{\rm inv}$ distortion in the PRF due to the Lorenz boost in the out direction resulting in values for $R_{\rm side}$ and $R_{\rm long}$ being smaller than $R_{\rm out}$.
For pions, the $\lambda$ values are also reduced due to long-lived resonances such as $\eta$ and $\eta^{'}$.

The decrease of the pion and kaon sphericity integrated radii with increasing pair transverse mass in pp collisions was observed at LHC energies before as an indication of collective behavior, originating from either initial- or final-state correlations. However, there was not even an approximate $m_{\rm T}$ scaling shown. The intriguing analogy with heavy-ion collisions was interpreted as a possibility of the collective behavior due to quark--gluon plasma formation~\cite{ALICE:2011kmy,Abelev:2012sq,CMS:2019fur,ATLAS:2017shk}.

The study presented here demonstrates a similar decrease of radii for both spherical and jet-like events. There is an indication of an approximate $m_{\rm T}$ scaling of the radii for pions and kaons in spherical events if the kinematic correction for these particle species is taken into account.
However, the fact that the decreasing $m_{\rm T}$ dependence and even the approximate $m_{\rm T}$ scaling are observed in all multiplicity intervals including the lowest ones suggests that this effect does not support the hypothesis of collective expansion of hot and dense matter as in heavy-ion collisions. A more realistic hypothesis is that of common emission conditions for pions and kaons in pp collisions for the multiplicity intervals studied in this work, similar to the simplified scenario described in Ref.~\cite{Kisiel:2014upa}.

\section{Summary}\label{Summary}
Correlations of two charged identical pions $\pi^\pm\pi^\pm$ and kaons K$^{\pm}$K$^{\pm}$ were measured in pp collisions at $\sqrt{s}=13$~TeV at the LHC.
One-dimensional pion and kaon correlations were classified using the global event-shape variable, transverse sphericity.
For spherical events, the correlation functions show a strong suppression of mini-jet contributions. In contrast, the correlation functions for jet-like events manifest large effects related to non-femtoscopic correlations at small relative momentum.

The Monte Carlo PYTHIA 8 (Monash) model describes both spherical and jet-like pion and kaon correlation functions outside the femtoscopic region effect. Therefore, it was used to subtract non-femtoscopic correlations from the femtoscopic signal region. The pion correlation functions corrected for mini-jet contributions as modeled by PYTHIA 8 are described by the exponential Bowler--Sinyukov function.
The Bowler--Sinyukov function using a single Gaussian was used to describe the kaon correlations. The pion femtoscopic radii extracted for spherical events are larger than those for jet-like events. Both pion and kaon radii demonstrate a decreasing trend with increasing pair transverse mass $m_{\rm T}$.

The Lorentz-boost corrected pion and kaon radii for both spherical and jet-like events show an approximate scaling behavior with $m_{\rm T}$ in all multiplicity intervals. The observation of such behavior for small multiplicity intervals, where the formation of QGP plasma seems to be impossible, can be interpreted as an indication of the emission occurring simultaneously for $\pi$ and K in pp collisions for all considered multiplicities.


\newenvironment{acknowledgement}{\relax}{\relax}
\begin{acknowledgement}
\section*{Acknowledgements}
\input{fa_2023-07-04_Opt_C.tex}    
\end{acknowledgement}

\bibliography{biblio_1}

\newpage
\appendix
\section{The ALICE Collaboration}
\label{app:collab}
\input{2023-07-04-Alice_Authorlist_2023-07-04_Opt_C.tex}  
\end{document}

%% file: fa_2023-07-04_Opt_C.tex

The ALICE Collaboration would like to thank all its engineers and technicians for their invaluable contributions to the construction of the experiment and the CERN accelerator teams for the outstanding performance of the LHC complex.
The ALICE Collaboration gratefully acknowledges the resources and support provided by all Grid centres and the Worldwide LHC Computing Grid (WLCG) collaboration.
The ALICE Collaboration acknowledges the following funding agencies for their support in building and running the ALICE detector:
A. I. Alikhanyan National Science Laboratory (Yerevan Physics Institute) Foundation (ANSL), State Committee of Science and World Federation of Scientists (WFS), Armenia;
Austrian Academy of Sciences, Austrian Science Fund (FWF): [M 2467-N36] and Nationalstiftung f\"{u}r Forschung, Technologie und Entwicklung, Austria;
Ministry of Communications and High Technologies, National Nuclear Research Center, Azerbaijan;
Conselho Nacional de Desenvolvimento Cient\'{\i}fico e Tecnol\'{o}gico (CNPq), Financiadora de Estudos e Projetos (Finep), Funda\c{c}\~{a}o de Amparo \`{a} Pesquisa do Estado de S\~{a}o Paulo (FAPESP) and Universidade Federal do Rio Grande do Sul (UFRGS), Brazil;
Bulgarian Ministry of Education and Science, within the National Roadmap for Research Infrastructures 2020-2027 (object CERN), Bulgaria;
Ministry of Education of China (MOEC) , Ministry of Science \& Technology of China (MSTC) and National Natural Science Foundation of China (NSFC), China;
Ministry of Science and Education and Croatian Science Foundation, Croatia;
Centro de Aplicaciones Tecnol\'{o}gicas y Desarrollo Nuclear (CEADEN), Cubaenerg\'{\i}a, Cuba;
Ministry of Education, Youth and Sports of the Czech Republic, Czech Republic;
The Danish Council for Independent Research | Natural Sciences, the VILLUM FONDEN and Danish National Research Foundation (DNRF), Denmark;
Helsinki Institute of Physics (HIP), Finland;
Commissariat \`{a} l'Energie Atomique (CEA) and Institut National de Physique Nucl\'{e}aire et de Physique des Particules (IN2P3) and Centre National de la Recherche Scientifique (CNRS), France;
Bundesministerium f\"{u}r Bildung und Forschung (BMBF) and GSI Helmholtzzentrum f\"{u}r Schwerionenforschung GmbH, Germany;
General Secretariat for Research and Technology, Ministry of Education, Research and Religions, Greece;
National Research, Development and Innovation Office, Hungary;
Department of Atomic Energy Government of India (DAE), Department of Science and Technology, Government of India (DST), University Grants Commission, Government of India (UGC) and Council of Scientific and Industrial Research (CSIR), India;
National Research and Innovation Agency - BRIN, Indonesia;
Istituto Nazionale di Fisica Nucleare (INFN), Italy;
Japanese Ministry of Education, Culture, Sports, Science and Technology (MEXT) and Japan Society for the Promotion of Science (JSPS) KAKENHI, Japan;
Consejo Nacional de Ciencia (CONACYT) y Tecnolog\'{i}a, through Fondo de Cooperaci\'{o}n Internacional en Ciencia y Tecnolog\'{i}a (FONCICYT) and Direcci\'{o}n General de Asuntos del Personal Academico (DGAPA), Mexico;
Nederlandse Organisatie voor Wetenschappelijk Onderzoek (NWO), Netherlands;
The Research Council of Norway, Norway;
Commission on Science and Technology for Sustainable Development in the South (COMSATS), Pakistan;
Pontificia Universidad Cat\'{o}lica del Per\'{u}, Peru;
Ministry of Education and Science, National Science Centre and WUT ID-UB, Poland;
Korea Institute of Science and Technology Information and National Research Foundation of Korea (NRF), Republic of Korea;
Ministry of Education and Scientific Research, Institute of Atomic Physics, Ministry of Research and Innovation and Institute of Atomic Physics and University Politehnica of Bucharest, Romania;
Ministry of Education, Science, Research and Sport of the Slovak Republic, Slovakia;
National Research Foundation of South Africa, South Africa;
Swedish Research Council (VR) and Knut \& Alice Wallenberg Foundation (KAW), Sweden;
European Organization for Nuclear Research, Switzerland;
Suranaree University of Technology (SUT), National Science and Technology Development Agency (NSTDA) and National Science, Research and Innovation Fund (NSRF via PMU-B B05F650021), Thailand;
Turkish Energy, Nuclear and Mineral Research Agency (TENMAK), Turkey;
National Academy of  Sciences of Ukraine, Ukraine;
Science and Technology Facilities Council (STFC), United Kingdom;
National Science Foundation of the United States of America (NSF) and United States Department of Energy, Office of Nuclear Physics (DOE NP), United States of America.
In addition, individual groups or members have received support from:
European Research Council, Strong 2020 - Horizon 2020 (grant nos. 950692, 824093), European Union;
Academy of Finland (Center of Excellence in Quark Matter) (grant nos. 346327, 346328), Finland.

%% file: 2023-07-04-Alice_Authorlist_2023-07-04_Opt_C.tex
\begin{flushleft} 
\small

S.~Acharya\,\orcidlink{0000-0002-9213-5329}\,$^{\rm 128}$, 
D.~Adamov\'{a}\,\orcidlink{0000-0002-0504-7428}\,$^{\rm 87}$, 
G.~Aglieri Rinella\,\orcidlink{0000-0002-9611-3696}\,$^{\rm 33}$, 
M.~Agnello\,\orcidlink{0000-0002-0760-5075}\,$^{\rm 30}$, 
N.~Agrawal\,\orcidlink{0000-0003-0348-9836}\,$^{\rm 52}$, 
Z.~Ahammed\,\orcidlink{0000-0001-5241-7412}\,$^{\rm 136}$, 
S.~Ahmad\,\orcidlink{0000-0003-0497-5705}\,$^{\rm 16}$, 
S.U.~Ahn\,\orcidlink{0000-0001-8847-489X}\,$^{\rm 72}$, 
I.~Ahuja\,\orcidlink{0000-0002-4417-1392}\,$^{\rm 38}$, 
A.~Akindinov\,\orcidlink{0000-0002-7388-3022}\,$^{\rm 142}$
% ...

\section*{Affiliation Notes}

$^{\rm I}$ Deceased\\
$^{\rm II}$ Also at: Max-Planck-Institut fur Physik, Munich, Germany\\
$^{\rm III}$ Also at: Italian National Agency for New Technologies, Energy and Sustainable Economic Development (ENEA), Bologna, Italy\\
$^{\rm IV}$ Also at: Dipartimento DET del Politecnico di Torino, Turin, Italy\\
$^{\rm V}$ Also at: Department of Applied Physics, Aligarh Muslim University, Aligarh, India\\
$^{\rm VI}$ Also at: Institute of Theoretical Physics, University of Wroclaw, Poland\\
$^{\rm VII}$ Also at: An institution covered by a cooperation agreement with CERN\\

\section*{Collaboration Institutes}

$^{1}$ A.I. Alikhanyan National Science Laboratory (Yerevan Physics Institute) Foundation, Yerevan, Armenia\\
$^{2}$ AGH University of Krakow, Cracow, Poland\\
$^{3}$ Bogolyubov Institute for Theoretical Physics, National Academy of Sciences of Ukraine, Kiev, Ukraine\\
$^{4}$ Bose Institute, Department of Physics  and Centre for Astroparticle Physics and Space Science (CAPSS), Kolkata, India\\
$^{5}$ California Polytechnic State University, San Luis Obispo, California, United States\\
$^{6}$ Central China Normal University, Wuhan, China\\
$^{7}$ Centro de Aplicaciones Tecnol\'{o}gicas y Desarrollo Nuclear (CEADEN), Havana, Cuba\\
$^{8}$ Centro de Investigaci\'{o}n y de Estudios Avanzados (CINVESTAV), Mexico City and M\'{e}rida, Mexico\\
$^{9}$ Chicago State University, Chicago, Illinois, United States\\
$^{10}$ China Institute of Atomic Energy, Beijing, China\\
$^{11}$ China University of Geosciences, Wuhan, China\\
$^{12}$ Chungbuk National University, Cheongju, Republic of Korea\\
$^{13}$ Comenius University Bratislava, Faculty of Mathematics, Physics and Informatics, Bratislava, Slovak Republic\\
$^{14}$ COMSATS University Islamabad, Islamabad, Pakistan\\
$^{15}$ Creighton University, Omaha, Nebraska, United States\\
$^{16}$ Department of Physics, Aligarh Muslim University, Aligarh, India\\
$^{17}$ Department of Physics, Pusan National University, Pusan, Republic of Korea\\
$^{18}$ Department of Physics, Sejong University, Seoul, Republic of Korea\\
$^{19}$ Department of Physics, University of California, Berkeley, California, United States\\
$^{20}$ Department of Physics, University of Oslo, Oslo, Norway\\
$^{21}$ Department of Physics and Technology, University of Bergen, Bergen, Norway\\
$^{22}$ Dipartimento di Fisica, Universit\`{a} di Pavia, Pavia, Italy\\
$^{23}$ Dipartimento di Fisica dell'Universit\`{a} and Sezione INFN, Cagliari, Italy\\
$^{24}$ Dipartimento di Fisica dell'Universit\`{a} and Sezione INFN, Trieste, Italy\\
$^{25}$ Dipartimento di Fisica dell'Universit\`{a} and Sezione INFN, Turin, Italy\\
$^{26}$ Dipartimento di Fisica e Astronomia dell'Universit\`{a} and Sezione INFN, Bologna, Italy\\
$^{27}$ Dipartimento di Fisica e Astronomia dell'Universit\`{a} and Sezione INFN, Catania, Italy\\
$^{28}$ Dipartimento di Fisica e Astronomia dell'Universit\`{a} and Sezione INFN, Padova, Italy\\
$^{29}$ Dipartimento di Fisica `E.R.~Caianiello' dell'Universit\`{a} and Gruppo Collegato INFN, Salerno, Italy\\
$^{30}$ Dipartimento DISAT del Politecnico and Sezione INFN, Turin, Italy\\
$^{31}$ Dipartimento di Scienze MIFT, Universit\`{a} di Messina, Messina, Italy\\
$^{32}$ Dipartimento Interateneo di Fisica `M.~Merlin' and Sezione INFN, Bari, Italy\\
$^{33}$ European Organization for Nuclear Research (CERN), Geneva, Switzerland\\
$^{34}$ Faculty of Electrical Engineering, Mechanical Engineering and Naval Architecture, University of Split, Split, Croatia\\
$^{35}$ Faculty of Engineering and Science, Western Norway University of Applied Sciences, Bergen, Norway\\
$^{36}$ Faculty of Nuclear Sciences and Physical Engineering, Czech Technical University in Prague, Prague, Czech Republic\\
$^{37}$ Faculty of Physics, Sofia University, Sofia, Bulgaria\\
$^{38}$ Faculty of Science, P.J.~\v{S}af\'{a}rik University, Ko\v{s}ice, Slovak Republic\\
$^{39}$ Frankfurt Institute for Advanced Studies, Johann Wolfgang Goethe-Universit\"{a}t Frankfurt, Frankfurt, Germany\\
$^{40}$ Fudan University, Shanghai, China\\
$^{41}$ Gangneung-Wonju National University, Gangneung, Republic of Korea\\
$^{42}$ Gauhati University, Department of Physics, Guwahati, India\\
$^{43}$ Helmholtz-Institut f\"{u}r Strahlen- und Kernphysik, Rheinische Friedrich-Wilhelms-Universit\"{a}t Bonn, Bonn, Germany\\
$^{44}$ Helsinki Institute of Physics (HIP), Helsinki, Finland\\
$^{45}$ High Energy Physics Group,  Universidad Aut\'{o}noma de Puebla, Puebla, Mexico\\
$^{46}$ Horia Hulubei National Institute of Physics and Nuclear Engineering, Bucharest, Romania\\
$^{47}$ HUN-REN Wigner Research Centre for Physics, Budapest, Hungary\\
$^{48}$ Indian Institute of Technology Bombay (IIT), Mumbai, India\\
$^{49}$ Indian Institute of Technology Indore, Indore, India\\
$^{50}$ INFN, Laboratori Nazionali di Frascati, Frascati, Italy\\
$^{51}$ INFN, Sezione di Bari, Bari, Italy\\
$^{52}$ INFN, Sezione di Bologna, Bologna, Italy\\
$^{53}$ INFN, Sezione di Cagliari, Cagliari, Italy\\
$^{54}$ INFN, Sezione di Catania, Catania, Italy\\
$^{55}$ INFN, Sezione di Padova, Padova, Italy\\
$^{56}$ INFN, Sezione di Pavia, Pavia, Italy\\
$^{57}$ INFN, Sezione di Torino, Turin, Italy\\
$^{58}$ INFN, Sezione di Trieste, Trieste, Italy\\
$^{59}$ Inha University, Incheon, Republic of Korea\\
$^{60}$ Institute for Gravitational and Subatomic Physics (GRASP), Utrecht University/Nikhef, Utrecht, Netherlands\\
$^{61}$ Institute of Experimental Physics, Slovak Academy of Sciences, Ko\v{s}ice, Slovak Republic\\
$^{62}$ Institute of Physics, Homi Bhabha National Institute, Bhubaneswar, India\\
$^{63}$ Institute of Physics of the Czech Academy of Sciences, Prague, Czech Republic\\
$^{64}$ Institute of Space Science (ISS), Bucharest, Romania\\
$^{65}$ Institut f\"{u}r Kernphysik, Johann Wolfgang Goethe-Universit\"{a}t Frankfurt, Frankfurt, Germany\\
$^{66}$ Instituto de Ciencias Nucleares, Universidad Nacional Aut\'{o}noma de M\'{e}xico, Mexico City, Mexico\\
$^{67}$ Instituto de F\'{i}sica, Universidade Federal do Rio Grande do Sul (UFRGS), Porto Alegre, Brazil\\
$^{68}$ Instituto de F\'{\i}sica, Universidad Nacional Aut\'{o}noma de M\'{e}xico, Mexico City, Mexico\\
$^{69}$ iThemba LABS, National Research Foundation, Somerset West, South Africa\\
$^{70}$ Jeonbuk National University, Jeonju, Republic of Korea\\
$^{71}$ Johann-Wolfgang-Goethe Universit\"{a}t Frankfurt Institut f\"{u}r Informatik, Fachbereich Informatik und Mathematik, Frankfurt, Germany\\
$^{72}$ Korea Institute of Science and Technology Information, Daejeon, Republic of Korea\\
$^{73}$ KTO Karatay University, Konya, Turkey\\
$^{74}$ Laboratoire de Physique Subatomique et de Cosmologie, Universit\'{e} Grenoble-Alpes, CNRS-IN2P3, Grenoble, France\\
$^{75}$ Lawrence Berkeley National Laboratory, Berkeley, California, United States\\
$^{76}$ Lund University Department of Physics, Division of Particle Physics, Lund, Sweden\\
$^{77}$ Nagasaki Institute of Applied Science, Nagasaki, Japan\\
$^{78}$ Nara Women{'}s University (NWU), Nara, Japan\\
$^{79}$ National and Kapodistrian University of Athens, School of Science, Department of Physics , Athens, Greece\\
$^{80}$ National Centre for Nuclear Research, Warsaw, Poland\\
$^{81}$ National Institute of Science Education and Research, Homi Bhabha National Institute, Jatni, India\\
$^{82}$ National Nuclear Research Center, Baku, Azerbaijan\\
$^{83}$ National Research and Innovation Agency - BRIN, Jakarta, Indonesia\\
$^{84}$ Niels Bohr Institute, University of Copenhagen, Copenhagen, Denmark\\
$^{85}$ Nikhef, National institute for subatomic physics, Amsterdam, Netherlands\\
$^{86}$ Nuclear Physics Group, STFC Daresbury Laboratory, Daresbury, United Kingdom\\
$^{87}$ Nuclear Physics Institute of the Czech Academy of Sciences, Husinec-\v{R}e\v{z}, Czech Republic\\
$^{88}$ Oak Ridge National Laboratory, Oak Ridge, Tennessee, United States\\
$^{89}$ Ohio State University, Columbus, Ohio, United States\\
$^{90}$ Physics department, Faculty of science, University of Zagreb, Zagreb, Croatia\\
$^{91}$ Physics Department, Panjab University, Chandigarh, India\\
$^{92}$ Physics Department, University of Jammu, Jammu, India\\
$^{93}$ Physics Program and International Institute for Sustainability with Knotted Chiral Meta Matter (SKCM2), Hiroshima University, Hiroshima, Japan\\
$^{94}$ Physikalisches Institut, Eberhard-Karls-Universit\"{a}t T\"{u}bingen, T\"{u}bingen, Germany\\
$^{95}$ Physikalisches Institut, Ruprecht-Karls-Universit\"{a}t Heidelberg, Heidelberg, Germany\\
$^{96}$ Physik Department, Technische Universit\"{a}t M\"{u}nchen, Munich, Germany\\
$^{97}$ Politecnico di Bari and Sezione INFN, Bari, Italy\\
$^{98}$ Research Division and ExtreMe Matter Institute EMMI, GSI Helmholtzzentrum f\"ur Schwerionenforschung GmbH, Darmstadt, Germany\\
$^{99}$ Saga University, Saga, Japan\\
$^{100}$ Saha Institute of Nuclear Physics, Homi Bhabha National Institute, Kolkata, India\\
$^{101}$ School of Physics and Astronomy, University of Birmingham, Birmingham, United Kingdom\\
$^{102}$ Secci\'{o}n F\'{\i}sica, Departamento de Ciencias, Pontificia Universidad Cat\'{o}lica del Per\'{u}, Lima, Peru\\
$^{103}$ Stefan Meyer Institut f\"{u}r Subatomare Physik (SMI), Vienna, Austria\\
$^{104}$ SUBATECH, IMT Atlantique, Nantes Universit\'{e}, CNRS-IN2P3, Nantes, France\\
$^{105}$ Sungkyunkwan University, Suwon City, Republic of Korea\\
$^{106}$ Suranaree University of Technology, Nakhon Ratchasima, Thailand\\
$^{107}$ Technical University of Ko\v{s}ice, Ko\v{s}ice, Slovak Republic\\
$^{108}$ The Henryk Niewodniczanski Institute of Nuclear Physics, Polish Academy of Sciences, Cracow, Poland\\
$^{109}$ The University of Texas at Austin, Austin, Texas, United States\\
$^{110}$ Universidad Aut\'{o}noma de Sinaloa, Culiac\'{a}n, Mexico\\
$^{111}$ Universidade de S\~{a}o Paulo (USP), S\~{a}o Paulo, Brazil\\
$^{112}$ Universidade Estadual de Campinas (UNICAMP), Campinas, Brazil\\
$^{113}$ Universidade Federal do ABC, Santo Andre, Brazil\\
$^{114}$ Universitatea Nationala de Stiinta si Tehnologie Politehnica Bucuresti, Bucharest, Romania\\
$^{115}$ University of Cape Town, Cape Town, South Africa\\
$^{116}$ University of Derby, Derby, United Kingdom\\
$^{117}$ University of Houston, Houston, Texas, United States\\
$^{118}$ University of Jyv\"{a}skyl\"{a}, Jyv\"{a}skyl\"{a}, Finland\\
$^{119}$ University of Kansas, Lawrence, Kansas, United States\\
$^{120}$ University of Liverpool, Liverpool, United Kingdom\\
$^{121}$ University of Science and Technology of China, Hefei, China\\
$^{122}$ University of South-Eastern Norway, Kongsberg, Norway\\
$^{123}$ University of Tennessee, Knoxville, Tennessee, United States\\
$^{124}$ University of the Witwatersrand, Johannesburg, South Africa\\
$^{125}$ University of Tokyo, Tokyo, Japan\\
$^{126}$ University of Tsukuba, Tsukuba, Japan\\
$^{127}$ Universit\"{a}t M\"{u}nster, Institut f\"{u}r Kernphysik, M\"{u}nster, Germany\\
$^{128}$ Universit\'{e} Clermont Auvergne, CNRS/IN2P3, LPC, Clermont-Ferrand, France\\
$^{129}$ Universit\'{e} de Lyon, CNRS/IN2P3, Institut de Physique des 2 Infinis de Lyon, Lyon, France\\
$^{130}$ Universit\'{e} de Strasbourg, CNRS, IPHC UMR 7178, F-67000 Strasbourg, France, Strasbourg, France\\
$^{131}$ Universit\'{e} Paris-Saclay, Centre d'Etudes de Saclay (CEA), IRFU, D\'{e}partment de Physique Nucl\'{e}aire (DPhN), Saclay, France\\
$^{132}$ Universit\'{e}  Paris-Saclay, CNRS/IN2P3, IJCLab, Orsay, France\\
$^{133}$ Universit\`{a} degli Studi di Foggia, Foggia, Italy\\
$^{134}$ Universit\`{a} del Piemonte Orientale, Vercelli, Italy\\
$^{135}$ Universit\`{a} di Brescia, Brescia, Italy\\
$^{136}$ Variable Energy Cyclotron Centre, Homi Bhabha National Institute, Kolkata, India\\
$^{137}$ Warsaw University of Technology, Warsaw, Poland\\
$^{138}$ Wayne State University, Detroit, Michigan, United States\\
$^{139}$ Yale University, New Haven, Connecticut, United States\\
$^{140}$ Yonsei University, Seoul, Republic of Korea\\
$^{141}$  Zentrum  f\"{u}r Technologie und Transfer (ZTT), Worms, Germany\\
$^{142}$ Affiliated with an institute covered by a cooperation agreement with CERN\\
$^{143}$ Affiliated with an international laboratory covered by a cooperation agreement with CERN.\\

\end{flushleft} 